\documentclass[aps,prb,twocolumn]{revtex4-1}

\usepackage[pdftex]{graphics}
\usepackage{amssymb,amsmath}

\begin{document}

\title{Asymmetric recombination and electron spin relaxation in the semiclassical theory of radical pair reactions}
\author{Alan M.~Lewis}
\affiliation{Department of Chemistry, University of Oxford, Physical and Theoretical Chemistry Laboratory, South Parks Road, Oxford, OX1 3QZ, UK}
\author{David E. Manolopoulos}
\affiliation{Department of Chemistry, University of Oxford, Physical and Theoretical Chemistry Laboratory, South Parks Road, Oxford, OX1 3QZ, UK}
\author{P. J. Hore}
\affiliation{Department of Chemistry, University of Oxford, Physical and Theoretical Chemistry Laboratory, South Parks Road, Oxford, OX1 3QZ, UK}

\begin{abstract}
We describe how the semiclassical theory of radical pair recombination reactions recently introduced by two of us  [D. E. Manolopoulos and P.~J.~Hore, J. Chem. Phys. {\bf 139}, 124106 (2013)] can be generalised to allow for different singlet and triplet recombination rates. This is a non-trivial generalisation because when the recombination rates are different the recombination process is dynamically coupled to the coherent electron spin dynamics of the radical pair. Furthermore, because the recombination operator is a two-electron operator, it is no longer sufficient simply to consider the two electrons as classical vectors: one has to consider the complete set of 16 two-electron spin operators as independent classical variables. The resulting semiclassical theory is first validated by comparison with exact quantum mechanical results for a model radical pair containing 12 nuclear spins. It is then used to shed light on the spin dynamics of a carotenoid-porphyrin-fullerene (CPF) triad containing considerably more nuclear spins which has recently been used to establish a \lq proof of principle' for the operation of a chemical compass [K. Maeda {\em et al.}, Nature {\bf 453}, 387 (2008)]. We find in particular that the intriguing biphasic behaviour that has been observed in the effect of an Earth-strength magnetic field on the time-dependent survival probability of the photo-excited C$^{\cdot +}$PF$^{\cdot -}$ radical pair arises from a delicate balance between its asymmetric recombination and the relaxation of the electron spin in the carotenoid radical.
\end{abstract}

\maketitle

\section{Introduction}

Conservation of spin angular momentum has profound consequences for the chemistry of transient paramagnetic molecules because it allows magnetic interactions that are much weaker than thermal energies to influence reaction rates and product yields. Observable phenomena include magnetic field effects\cite{Rodgers09,Hore12,Volk95} and magnetic isotope effects\cite{Buchachenko09} on chemical reactivity, and the production of large electron\cite{Forbes13,Mobius09} and nuclear\cite{Lee14,Goez09} spin hyperpolarizations -- effects that collectively constitute the field of spin chemistry.\cite{Steiner89} The most intensively studied entities that exhibit these behaviours are the radical pairs that are formed as transient intermediates in thermal and photochemical reactions in condensed phases. 

Provided spin-orbit coupling is weak, pairs of radicals are created, for example by bond fission or electron transfer, in the same overall electronic spin state (singlet or triplet) as their chemical precursors. However, the spin eigenstates of radical pairs are normally neither singlet nor triplet because of the symmetry-breaking that arises from electron-nuclear hyperfine and/or Zeeman interactions. As a result, radical pairs form in superposition states which evolve coherently at frequencies determined by the internal and external magnetic interactions of the radicals. These interactions determine the time-dependent probability that a radical pair is singlet or triplet and therefore influence both the yields of spin-selective reactions and the lifetime of the radical pair itself if the competing singlet and triplet reaction pathways have different rate constants. Radical recombination and spin relaxation can be slow enough to allow even an Earth-strength magnetic field ($\sim 50$ $\mu$T) to have a measurable effect via the electron Zeeman interaction ($\sim 10^{-7}$ $k_{\rm B}T$ at 300 K).\cite{Maeda08}

The theory of the radical pair mechanism is well developed and in principle capable of exact quantitative interpretation of a wide variety of experimental data.\cite{Steiner89} In general one needs to treat the time-dependent quantum mechanics of the electron-nuclear spin system, including the relevant spin interactions (Zeeman, hyperfine, dipolar, exchange, quadrupolar, $\ldots$), chemical reaction steps, spin relaxation pathways, molecular motions, and so on. A major difficulty encountered in such calculations, however, is that they scale exponentially with the number of nuclear spins, $N$, and it is rarely possible to perform exact quantum simulations for $N$ greater than about 10. Unfortunately, organic radicals often have more than 10 nuclei with significant hyperfine couplings, especially when the molecular orbital that contains the unpaired electron is delocalised as in the case of extended $\pi$-systems. The problem becomes particularly acute for biological radical pairs, for example those formed naturally in photosynthetic reaction centre proteins during the initial steps of solar energy conversion\cite{Volk95,Alia13} and those thought to act as the magnetic compass sensors in cryptochrome proteins in the avian retina.\cite{Rodgers09b,Mouritsen12,Dodson13}

A potential solution to the exponential scaling problem was put forward by Schulten and Wolynes (SW) in the early days of spin chemistry.\cite{Schulten78} They modelled the electron spin motion induced by hyperfine coupling to many nuclear spins by means of a semiclassical (SC) approximation in which the electron spin precesses around a hyperfine-weighted sum of nuclear spin vectors, assumed to be fixed in space. Although successful in some circumstances,\cite{Schulten78} this approach conflicts with Newton's third law of motion, because the torque exerted on the electron spin by the nuclear spins is not balanced by an equal and opposite torque on the nuclear spins from the electron. Our recent extension to the SW theory\cite{Manolopoulos13} eliminates this inconsistency by allowing each individual nuclear spin in the radical to precess around the electron spin. The resulting SC theory approaches quantitative agreement with quantum mechanics as $N$ increases, reproduces the correct behaviour of long-lived radical pairs in applied magnetic fields, scales linearly with $N$ and so is applicable to arbitrarily large radicals, and is faster than the exact quantum mechanical (QM) calculation for $N \gtrsim 12$.\cite{Manolopoulos13}

Both SC approaches\cite{Schulten78,Manolopoulos13} are applicable to the case in which the rate constants for the spin-selective reactions of singlet and triplet radical pairs are identical. This has been a popular simplifying assumption in the past,\cite{Timmel98,Gauger11,Lee14b} but it is rarely realistic. For solution-phase reactions, for example, singlet radical pairs often react to form diamagnetic products in a diffusion-controlled manner while the corresponding triplets are unreactive because of the absence of low-energy triplet product states. Even when spin-allowed singlet and triplet reaction routes coexist, their rate constants are rarely identical, for example because singlet and triplet electron transfers have different thermodynamic driving forces.\cite{Volk95}

Here we generalise our earlier SC theory\cite{Manolopoulos13} to allow for asymmetric recombination. To do this, we consider the complete set of 16 two-electron spin operators of the radical pair as independent classical variables. Having validated this approach against the exact quantum mechanical results for a model radical pair with $N=12$ spin-$1/2$  nuclei, we use it to account for the experimentally observed effects of Earth-strength magnetic fields (39 and 49 $\mu$T) on the lifetime of a radical pair containing $N=45$ protons.\cite{Maeda08}

\section{Theory}

\subsection{Quantum mechanics}

\begin{figure}[t]
\centering
\resizebox{0.5\columnwidth}{!} {\includegraphics{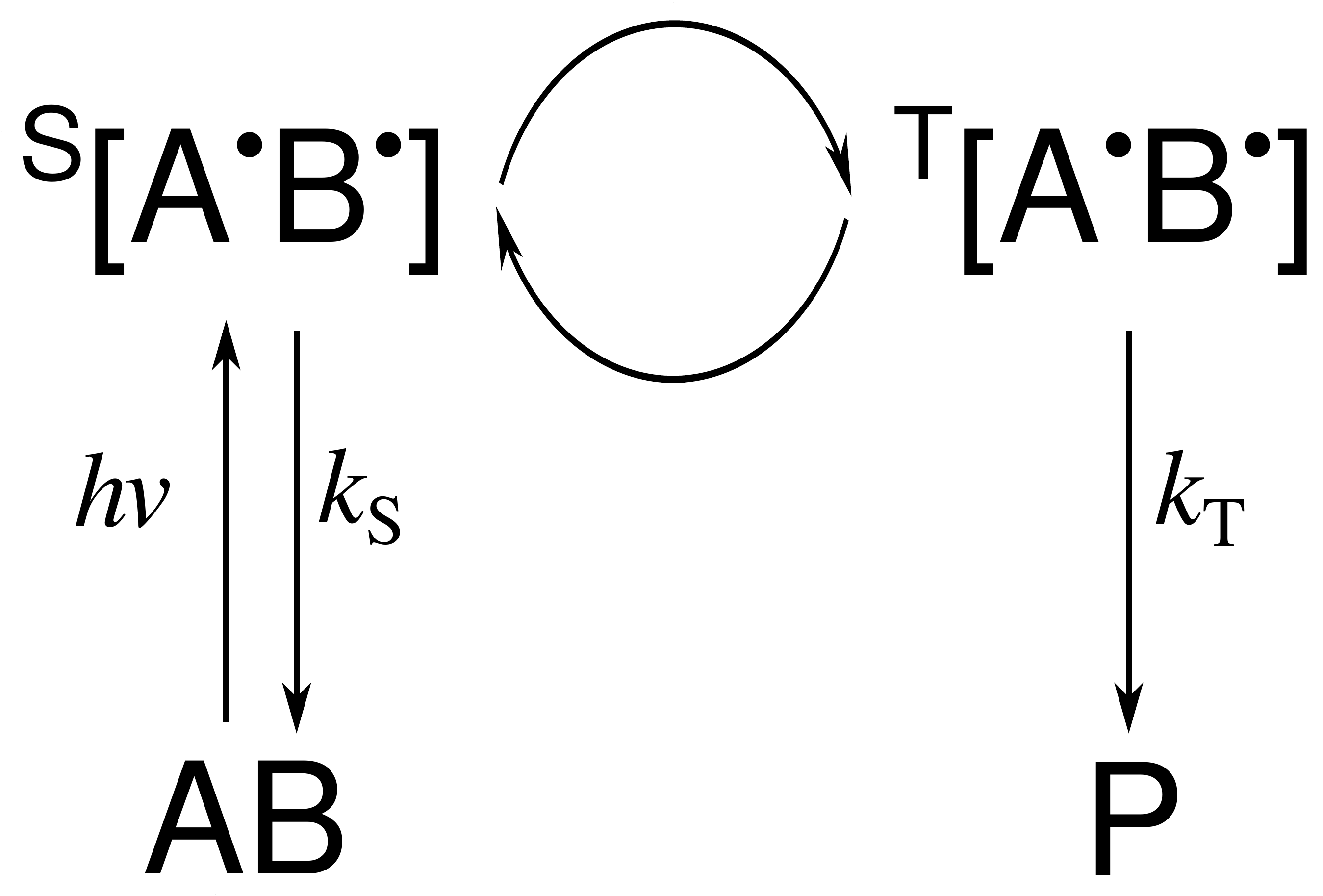}}
\caption{An idealised radical pair recombination reaction. The product P may or may not be the same as AB.}
\end{figure}

Consider the idealised radical pair recombination reaction illustrated in Fig.~1. The radical pair is formed photochemically in its singlet state at time $t=0$, and then undergoes coherent electron spin evolution between the singlet and triplet states while the singlet component recombines with a first order rate constant $k_{\rm S}$ and the triplet component recombines with a rate constant $k_{\rm T}$. 

Assuming that the dominant interactions in the radical pair are the Zeeman interactions between the electron spins and an applied magnetic field and the hyperfine interactions between the electron and nuclear spins within each radical, the Hamiltonian that governs the coherent electron spin evolution is\cite{Schulten78}
$$
\hat{H} = \hat{H}_1+\hat{H}_2, \eqno(1)
$$
where
$$
\hat{H}_i = \boldsymbol{\omega}_i\cdot\hat{\bf S}_i+\sum_{k=1}^{N_i} a_{ik}\,\hat{\bf I}_{ik}\cdot\hat{\bf S}_i. \eqno(2)
$$
Here $\boldsymbol{\omega}_i = -\gamma_i{\bf B}$, where $\gamma_i$ is the gyromagnetic ratio of the electron in radical $i$ and ${\bf B}$ is the applied magnetic field, $a_{ik}$ is the hyperfine coupling constant between the $k$th nuclear spin in this radical and the electron spin, $\hat{\bf I}_{ik}$ and $\hat{\bf S}_i$ are the corresponding electron and nuclear spin operators, and $N_i$ is the number of nuclear spins in radical $i$. As is conventional in this field, we shall work throughout this paper in a unit system in which $\hbar=1$, where the unit of time is the reciprocal of the unit that is used to specify $\boldsymbol{\omega}_i$ and $a_{ik}$.

The Hamiltonian in Eq.~(1) does not couple the electron spins because $\hat{H}_1$ and $\hat{H}_2$ operate on each radical separately: $[\hat{H}_1,\hat{H}_2]=0$. However, the recombination operator that governs the singlet and triplet recombination processes generally does couple the two electron spins. This recombination operator is\cite{Haberkorn76}
$$
\hat{K} = {k_{\rm S}\over 2}\hat{P}_{\rm S}+{k_{\rm T}\over 2}\hat{P}_{\rm T}, \eqno(3)
$$
where
$$
\hat{P}_{\rm S} = {1\over 4}\hat{\bf 1}-\hat{\bf S}_1\cdot\hat{\bf S}_2, \eqno(4)
$$
and
$$
\hat{P}_{\rm T} = {3\over 4}\hat{\bf 1}+\hat{\bf S}_1\cdot\hat{\bf S}_2, \eqno(5)
$$
are the projection operators onto the singlet and triplet electronic subspaces. When $k_{\rm S}$ and $k_{\rm T}$ are the same, $k_{\rm S}=k_{\rm T}=\bar{k}$, the recombination operator is simply $\hat{K}=(\bar{k}/2)\hat{\bf 1}$, a multiple of the unit operator. But in the general case where $k_{\rm S}\not=k_{\rm T}$, the operator $\hat{K}$ involves $\hat{\bf S}_1\cdot\hat{\bf S}_2$, which couples the two electron spins. 

Since the radical pair is formed in the singlet state, its initial density operator is
$$
\hat{\rho}(0) = {1\over Z_1Z_2}\hat{P}_{\rm S}, \eqno(6)
$$
where $Z_i=\prod_{k=1}^{N_i} (2I_{ik}+1)$ is the number of nuclear spin states in radical $i$. This density operator evolves in time according to
$$
{d\over dt}\hat{\rho}(t) = -i[\hat{H},\hat{\rho}(t)]-\{\hat{K},\hat{\rho}(t)\}, \eqno(7)
$$
where $[\hat{A},\hat{B}]$ is the commutator and $\{\hat{A},\hat{B}\}$ the anticommutator of $\hat{A}$ and $\hat{B}$; the solution of Eq.~(7) is
$$
\hat{\rho}(t) = e^{-i\hat{H}t-\hat{K}t}\,\hat{\rho}(0)\,e^{+i\hat{H}t-\hat{K}t}. \eqno(8)
$$
This can be used to calculate the expectation value of an observable $A$ represented by an operator $\hat{A}$ at time $t$ as
$$
A(t) = {\rm tr}\left[\hat{\rho}(t)\hat{A}(0)\right], \eqno(9)
$$
where the trace is over all $4Z_1Z_2$ states in the combined electronic and nuclear spin Hilbert space of the radical pair. Equation~(9) can be written equivalently as
$$
A(t) = {\rm tr}\left[\hat{\rho}(0)\hat{A}(t)\right], \eqno(10)
$$
where
$$
\hat{A}(t) = e^{+i\hat{H}t-\hat{K}t}\,\hat{A}(0)\,e^{-i\hat{H}t-\hat{K}t} \eqno(11)
$$
satisfies the Heisenberg equation of motion
$$
{d\over dt}\hat{A}(t) = +i[\hat{H},\hat{A}(t)]-\{\hat{K},\hat{A}(t)\} \eqno(12)
$$
subject to the initial condition $\hat{A}(0)=\hat{A}$. This Heisenberg picture will prove to be especially useful in the development of the semiclassical approximation for asymmetric recombination described below.

The primary experimental observables are the time-dependent singlet and triplet probabilities of the radical pair. In light of Eq.~(10), these are given by
$$
P_{\rm S}(t) = {\rm tr}\left[\hat{\rho}(0)\hat{P}_{\rm S}(t)\right] 
=  {1\over Z_1Z_2}{\rm tr}\left[\hat{P}_{\rm S}(0)\hat{P}_{\rm S}(t)\right], 
\eqno(13)
$$
and
$$
P_{\rm T}(t) = {\rm tr}\left[\hat{\rho}(0)\hat{P}_{\rm T}(t)\right]
= {1\over Z_1Z_2}{\rm tr}\left[\hat{P}_{\rm S}(0)\hat{P}_{\rm T}(t)\right], 
\eqno(14)
$$
where the operators $\hat{P}_{\rm S}$ and $\hat{P}_{\rm T}$ are defined in Eqs.~(4) and~(5) and the dynamics that takes these operators from time 0 to time $t$ is that in Eqs.~(11) and~(12). Secondary observables include the singlet and triplet quantum yields of the reaction. These are given in terms of $P_{\rm S}(t)$ and $P_{\rm T}(t)$ by
$$
\Phi_{\rm S} = \int_0^{\infty} k_{\rm S}P_{\rm S}(t)\,dt, \eqno(15)
$$
and
$$
\Phi_{\rm T} = \int_0^{\infty} k_{\rm T}P_{\rm T}(t)\,dt, \eqno(16)
$$
which sum correctly to give
$$
\Phi_{\rm S}+\Phi_{\rm T} = {\rm tr}\left\{\hat{\rho}(0)\int_0^{\infty}\left(k_{\rm S}\hat{P}_{\rm S}(t)+k_{\rm T}\hat{P}_{\rm T}(t)\right)dt\right\}
\phantom{xx}
$$
$$
={\rm tr}\left\{\hat{\rho}(0) \int_0^{\infty} 2\hat{K}(t)\,dt\right\}
\phantom{xxxxxxx}
$$
$$
\phantom{xxxxx}
={\rm tr}\left\{\hat{\rho}(0) \int_0^{\infty} e^{+i\hat{H}t-\hat{K}t}\,2\hat{K}\,e^{-i\hat{H}t-\hat{K}t}\,dt\right\}
$$
$$
\phantom{xxxxxxx.}
={\rm tr}\left\{\hat{\rho}(0) \int_0^{\infty} {d\over dt}\left(-e^{+i\hat{H}t-\hat{K}t}e^{-i\hat{H}t-\hat{K}t}\right)\,dt\right\}
$$
$$
\phantom{.}
={\rm tr}\left\{\hat{\rho}(0) \left[-e^{+i\hat{H}t-\hat{K}t}e^{-i\hat{H}t-\hat{K}t}\right]_0^{\infty}\right\}
$$
$$
={\rm tr}\left\{\hat{\rho}(0)\,\hat{\bf 1}\right\} = {\rm tr}\left\{\hat{\rho}(0)\right\} = 1.
\phantom{xxxx.} \eqno(17)
$$

For a general asymmetric radical pair recombination reaction, this is as far as the development can be taken: one is obliged to solve a time-dependent quantum mechanical problem in a Hilbert space of $4Z_1Z_2$ states. However, in the special case where $k_{\rm S}=k_{\rm T}=\bar{k}$, there is a significant simplification: because $\hat{K}=(\bar{k}/2)\hat{\bf 1}$ commutes with everything, the recombination can be factored out of the dynamics, which allows Eq.~(13) (for example) to be re-written as
$$
P_{\rm S}(t) = e^{-\bar{k}t}P_{\rm S}^0(t), \eqno(18)
$$ 
where $P_{\rm S}^0(t)$ is the singlet probability in the absence of any recombination:
$$
P_{\rm S}^0(t) = {1\over Z_1Z_2}{\rm tr}\left[\hat{P}_{\rm S}\,e^{+i\hat{H}t}\,\hat{P}_{\rm S}\,e^{-i\hat{H}t}\right]. \eqno(19)
$$
Furthermore, since $\hat{H}=\hat{H}_1+\hat{H}_2$ and $[\hat{H}_1,\hat{H}_2]=0$, the dynamics of the two radicals are no longer coupled. In fact, Eq.~(19) can be rearranged into the form\cite{Schulten78}
$$
P_{\rm S}^{0}(t) = {1\over 4}+\sum_{\alpha\beta}R_{\alpha\beta}^{(1)}(t)R_{\alpha\beta}^{(2)}(t), \eqno(20)
$$
where 
$$
R_{\alpha\beta}^{(i)}(t) = {1\over Z_i}{\rm tr}_i\left[\hat{S}_{i\alpha}\,e^{+i\hat{H}_it}\,\hat{S}_{i\beta}\,e^{-i\hat{H}_it}\right]
$$
$$
\equiv {1\over Z_i}{\rm tr}_i\left[\hat{S}_{i\alpha}(0)\,\hat{S}_{i\beta}(t)\right] \eqno(21)
$$
is an electron spin correlation tensor for the spin on radical $i$ in the absence of any recombination and ${\rm tr}_i$ denotes a trace over the $2Z_i$ states in the combined electronic and nuclear spin Hilbert space of this radical. The problem of calculating the time-dependent singlet and triplet probabilities of the radical pair is thus reduced in the symmetric recombination case to one of calculating the electron spin correlation tensors of each radical separately.\cite{Schulten78}

\subsection{Semiclassical approximation} 

The semiclassical theory introduced in Ref.~\cite{Manolopoulos13} was restricted to the symmetric recombination case and consisted of the following classical vector model approximation to the electron spin correlation tensor in Eq.~(21):
$$
R_{\alpha\beta}^{(i)}(t) \simeq {2\over 4\pi}\int d\Omega_{S_i} \prod_{k=1}^{N_i} {1\over 4\pi} \int d\Omega_{I_{ik}} S_{i\alpha}(0)S_{i\beta}(t). \eqno(22)
$$
Here the trace over the electron spin states has been replaced by an integral over the orientation $\Omega_{S_i}$ of a classical electron spin vector of length $\sqrt{S_i(S_i+1)}$,
$$
{\rm tr}_{S_i} \to {2S_i+1\over 4\pi} \int d\Omega_{S_i}, \eqno(23)
$$
and the trace over the spin states of the $k$-th nucleus in the radical has been replaced by an integral over the orientation $\Omega_{I_{ik}}$ of a classical nuclear spin vector of length $\sqrt{I_{ik}(I_{ik}+1)}$,
$$
{\rm tr}_{I_{ik}} \to {2I_{ik}+1\over 4\pi} \int d\Omega_{I_{ik}}. \eqno(24)
$$
The factor of $2$ in Eq.~(22) arises because $(2S_i+1)=2$, and there are no factors of $(2I_{ik}+1)$ because these combine to cancel the $1/Z_i$ pre-factor in Eq.~(21). For each initial electronic and nuclear spin orientation that is integrated over in Eq.~(22), the correlated product $S_{i\alpha}(0)S_{i\beta}(t)$ is obtained from the coupled classical spin dynamics,\cite{Manolopoulos13}
$$
{d\over dt}{\bf I}_{ik} = a_{ik}\,{\bf S}_i \wedge {\bf I}_{ik}, \eqno(25)
$$
and
$$
{d\over dt}{\bf S}_i = \left[\boldsymbol{\omega}_i+\sum_{k=1}^{N_i} a_{ik}\,{\bf I}_{ik}\right] \wedge {\bf S}_i, \eqno(26)
$$
where the wedges ($\wedge$) denote vector products.

Despite its simplicity, this semiclassical approximation actually gives the exact quantum mechanical electron spin correlation tensor of a radical with no nuclear spins.\cite{Manolopoulos13} It misses the coherent quantum oscillations in the spin correlation tensor of a radical with just one nuclear spin, but it is found numerically to give a tensor that becomes increasingly accurate as the number of nuclear spins increases and the environment of the electron spin becomes more complex, and there are good reasons to believe that it will again become exact in the limit where the number of nuclear spins in the radical tends to infinity.\cite{Manolopoulos13,Erlingsson04,Chen07}

The goal is thus to generalise this theory to the case of asymmetric recombination, in which one has to work in the combined Hilbert space of the radical pair and the correlation functions of interest have the form of $P_{\rm S}(t)$ in Eq.~(13). In order to do this, let us suppose from the outset that we are going to continue to treat the nuclear spins as classical vectors, so as to obtain a theory that is applicable to arbitrarily large radicals. This can be accomplished by replacing every occurrence of the vector operator ${\hat{\bf I}}_{ik}$ in the Heisenberg equations of motion in Eq.~(12) with a classical nuclear spin angular momentum vector ${\bf I}_{ik}$. So what happens when we attempt to do this without making any further approximation?

To answer this question, we need to know the complete set of operators representing the dynamical variables of the two electron spins. Since an $n\times n$ Hermitian density matrix is specified by $n(n+1)/2+n(n-1)/2=n^2$ independent parameters (the real and imaginary parts of its upper triangular matrix elements), there must be $4^2=16$ independent Hermitian two-electron spin operators, which are readily identified to be $\hat{\bf S}_{1}$ (three operators), $\hat{\bf S}_{2}$ (another three), $\hat{\bf T}_{12}=\hat{\bf S}_1\otimes\hat{\bf S}_2$ (another nine), and $\hat{\bf 1}$ (the last one). 

The Heisenberg equations of motion for these electron spin operators and the nuclear spin operators $\hat{\bf I}_{ik}$ are those in Eq.~(12). Closing these equations of motion in a naive way by replacing every occurrence of $\hat{\bf I}_{ik}$  with the classical vector ${\bf I}_{ik}$, and then taking the hats off the electron spin operators to convert them into classical variables, we find that 
$$
{d\over dt}{\bf I}_{ik} = a_{ik}{\bf S}_i\wedge {\bf I}_{ik}-\bar{k}\,{\bf I}_{ik}+4\Delta k\,{\rm tr}[{\bf T}_{12}]{\bf I}_{ik}, \eqno(27)
$$
$$
{d\over dt}{\bf S}_1 = \bar{\boldsymbol{\omega}}_1\wedge {\bf S}_1-\bar{k}\,{\bf S}_1+\Delta k\,{\bf S}_2, \eqno(28)
$$
$$
{d\over dt}{\bf S}_2 = \bar{\boldsymbol{\omega}}_2\wedge {\bf S}_2-\bar{k}\,{\bf S}_2+\Delta k\,{\bf S}_1, \eqno(29)
$$
$$
{d\over dt}{\bf T}_{12} = \bar{\boldsymbol{\omega}}_1\wedge {\bf T}_{12}-{\bf T}_{12}\wedge\bar{\boldsymbol{\omega}}_2-\bar{k}{\bf T}_{12}+\Delta k\,{\bf T}_{12}^{\rm T}+\Delta k\,\bar{P}_{\rm S}\,{\bf I}, \eqno(30)
$$
and
$$
{d\over dt}{\bf 1} = -\bar{k}\,{\bf 1}+4\Delta k\,{\rm tr}[{\bf T}_{12}], \eqno(31)
$$
where $\bar{k}=(k_{\rm S}+3k_{\rm T})/4$, $\Delta k=(k_{\rm S}-k_{\rm T})/4$, and
$$
\bar{\boldsymbol{\omega}}_i = \boldsymbol{\omega}_i+\sum_{k=1}^{N_i} a_{ik}\,{\bf I}_{ik}. \eqno(32)
$$
In Eqs.~(27) and~(31), ${\rm tr}[{\bf T}_{12}]$ denotes the trace of the tensor ${\bf T}_{12}$, which is the classical variable corresponding to the operator $\hat{\bf S}_1\cdot\hat{\bf S}_2$. In Eq.~(30), $\bar{\boldsymbol{\omega}}_1\wedge {\bf T}_{12}$ denotes the vector product of $\bar{\boldsymbol{\omega}}_1$ with each column of ${\bf T}_{12}$, ${\bf T}_{12}\wedge\bar{\boldsymbol{\omega}}_2$ denotes the vector product of each row of ${\bf T}_{12}$ with $\bar{\boldsymbol{\omega}}_2$, ${\bf T}_{12}^{\rm T}$ is the transpose of ${\bf T}_{12}$, ${\bf I}$ is a $(3\times 3)$ unit matrix, and
$$
\bar{P}_{\rm S} = {1\over 4}{\bf 1}-{\rm tr}[{\bf T}_{12}] \eqno(33)
$$
is the classical variable corresponding to the singlet projection operator $\hat{P}_{\rm S}$ in Eq.~(4).

There are several interesting points to make about these equations. The first is that, in the absence of any recombination, where $\bar{k}=\Delta k=0$, Eqs.~(27) to~(29) reduce correctly to give the classical vector model dynamics of the separate radicals in Eqs.~(25) and~(26), and Eqs.~(30) and~(31) give ${\bf T}_{12}={\bf S}_1\otimes {\bf S}_2$ and ${\bf 1}=1$ for all $t$. However, when $\bar{k}\not=0$ and $\Delta k\not=0$, ${\bf 1}$ (the classical variable corresponding to the unit operator) changes with time, and while it is still true that ${\bf T}_{12}={\bf S}_1\otimes{\bf S}_2$ at time $t=0$ this is no longer true at later times. This explains why all 16 electron spin operators have to be treated as independent classical variables when recombination is included in the equations of motion.

A more intriguing observation is that, in the case of symmetric recombination, where $\Delta k=0$ but $\bar{k}\not=0$, Eqs.~(27) to~(31) are inconsistent with our earlier semiclassical theory in Eqs.~(18) to~(26). In that theory, the electron spin in radical $i$ precesses around $N_i$ nuclear spin vectors of length $\sqrt{I_{ik}(I_{ik}+1)}$, each of which precesses in response around an electron spin vector of length $\sqrt{S_i(S_i+1)}$. In Eqs.~(27) to (29), the lengths of the nuclear and electron spin vectors decay with time, leading to a slowing down of the precessional motion. This is clearly unphysical. From the point of view of the classical vector model, either the radical pair has recombined, or it has not. If it has recombined, the precession is non-existent, and if it has not, there is no reason why the combined electronic and nuclear spin precession should slow down with time.

Since the only approximation we have made is to treat the nuclear spin operator $\hat{\bf I}_{ik}$ as a classical vector, and to use this to close the Heisenberg equations of motion, this must be at the root of the problem. In other words, the key equation that needs to be modified is Eq.~(27). It would be inappropriate to modify Eqs.~(28) and~(29), because they give the exact quantum mechanical result for the electron spin dynamics of a radical pair without any nuclear spins (see below). Moreover there is only one way to modify Eq.~(27) to restore consistency with our earlier semiclassical theory, which is to replace it with 
$$
{d\over dt}{\bf I}_{ik} = a_{ik}{\sqrt{S_i(S_i+1)}\over |{\bf S}_i|}{\bf S}_i\wedge {\bf I}_{ik}. \eqno(34)
$$
With this modification, each nuclear spin continues to precess around an electron spin vector of length $\sqrt{S_i(S_i+1)}$, and since there are no decay terms in Eq.~(34), the length of the nuclear spin vector ${\bf I}_{ik}$ remains fixed at $\sqrt{I_{ik}(I_{ik}+1)}$ for all time, so the electron spin precession does not slow down either.

To complete the specification of the approximation, we still need to specify the initial values of the nuclear and electron spin variables. For this, we use the classical vector model prescriptions in Eqs.~(23) and~(24) for the initial values of ${\bf S}_i$ and ${\bf I}_{ik}$, and set ${\bf T}_{12}={\bf S}_1\otimes{\bf S}_2$ and ${\bf 1}=1$ at time $t=0$. Our semiclassical approximation to the singlet probability $P_{\rm S}(t)$ in Eq.~(13) is thus
$$
P_{\rm S}(t) \simeq \prod_{i=1}^2 {2\over 4\pi}\int d\Omega_{S_i}\prod_{k=1}^{N_i}{1\over 4\pi} \int d\Omega_{I_{ik}}\,\bar{P}_{\rm S}(0)\bar{P}_{\rm S}(t), \eqno(35)
$$
where $\bar{P}_{\rm S}$ is defined in Eq.~(33), and the corresponding approximation to the triplet probability is
$$
P_{\rm T}(t) \simeq \prod_{i=1}^2 {2\over 4\pi}\int d\Omega_{S_i}\prod_{k=1}^{N_i}{1\over 4\pi} \int d\Omega_{I_{ik}}\,\bar{P}_{\rm S}(0)\bar{P}_{\rm T}(t), \eqno(36)
$$
where
$$
\bar{P}_{\rm T} = {3\over 4}{\bf 1}+{\rm tr}[{\bf T}_{12}]. \eqno(37)
$$

\subsection{An exact limit}

We have claimed above that the only approximation we have made in obtaining these equations is to treat the nuclear spins as classical vectors: the coherent quantum mechanical electron spin dynamics of the radical pair has been treated exactly. This can be demonstrated by regarding $\bar{\boldsymbol{\omega}}_1$ and $\bar{\boldsymbol{\omega}}_2$ in Eqs.~(28) to~(30) as constant vectors, as they are in a radical pair without any nuclear spins, and by eliminating integrals over the nuclear spin orientations in Eqs.~(35) and~(36). It is easy to show that the resulting equations give the correct results $P_{\rm S}(0)=1$ and $P_{\rm T}(0)=0$ at time $t=0$, simply by doing the integrals over $\Omega_{S_1}$ and $\Omega_{S_2}$. It is harder to show that they are correct at longer times, so we shall simply illustrate this here with a numerical example.\cite{Footnote1} Fig.~2 compares the quantum mechanical $P_{\rm S}(t)$ of Eq.~(13) and $P_{\rm T}(t)$ of Eq.~(14) with the semiclassical approximations in Eqs.~(35) and~(36) for a broken symmetry test case involving electron spin evolution and singlet and triplet recombination in the presence of fixed field vectors $\bar{\boldsymbol{\omega}}_1\not=\bar{\boldsymbol{\omega}}_2$. The quantum mechanical and semiclassical results are identical to graphical accuracy, and we have verified that the same is true for a number of other test cases both with and without $\bar{\boldsymbol{\omega}}_1=\bar{\boldsymbol{\omega}}_2$ and $k_{\rm S}=k_{\rm T}$ symmetry.

\begin{figure}[t]
\centering
\resizebox{0.8\columnwidth}{!} {\includegraphics{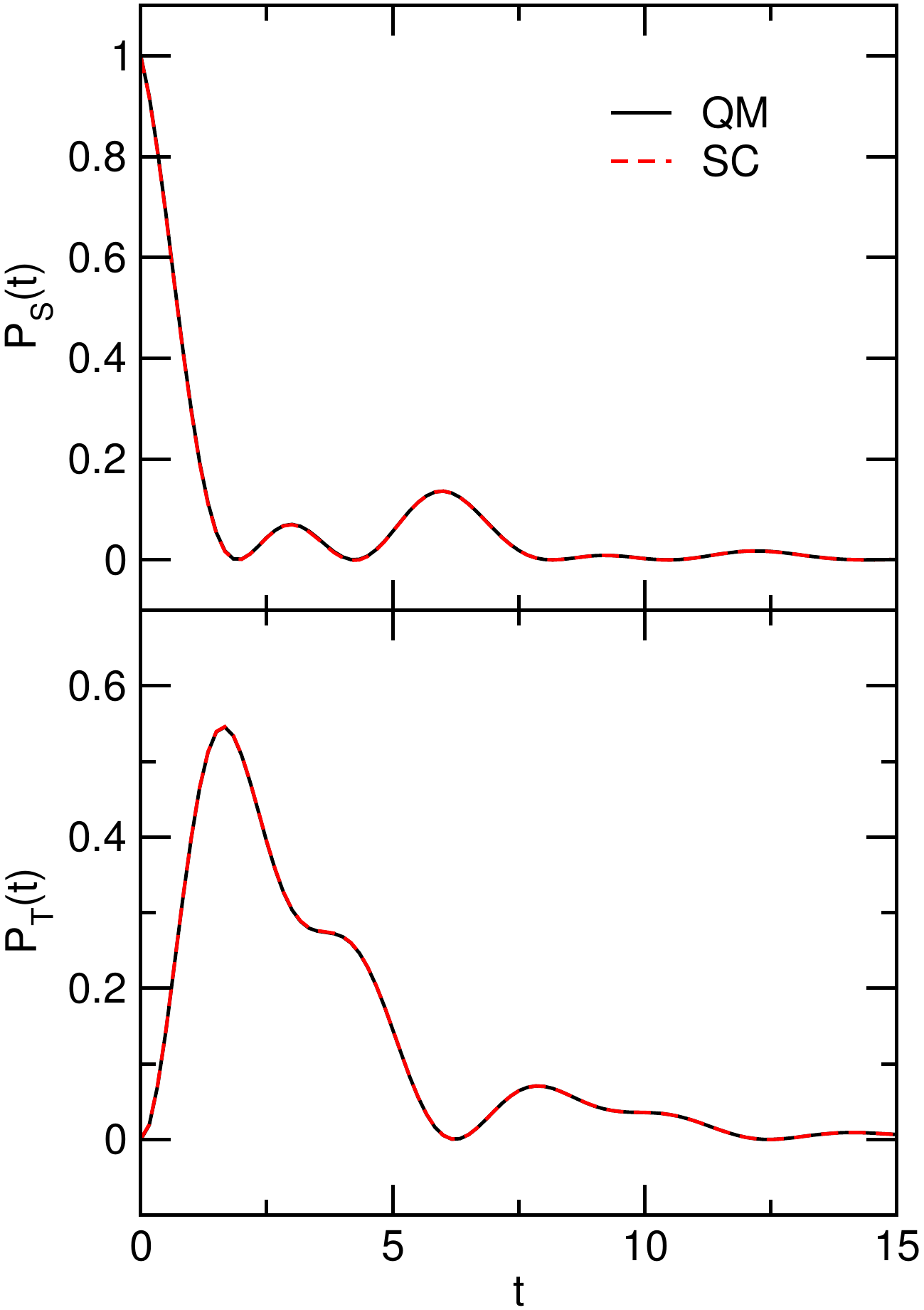}}
\caption{Comparison of quantum mechanical (QM) and semiclassical (SC) time-dependent singlet and triplet probabilities for a radical pair with $\bar{\boldsymbol{\omega}}_1=(0,0,1)$, $\bar{\boldsymbol{\omega}}_2=-(\sqrt{1/2},\sqrt{1/3},\sqrt{1/5})$, $k_{\rm S}=\sqrt{1/7}$, and $k_{\rm T}=\sqrt{1/11}$.}
\end{figure}

Fig.~2 is directly relevant to the present study because we shall be especially interested in how well the semiclassical theory performs for the calculation of two-electron observables such as $P_{\rm S}(t)$ and $P_{\rm T}(t)$. However, on closer inspection, one sees that the dynamics of the one-electron variables ${\bf S}_1$ and ${\bf S}_2$ in Eqs.~(28) and (29) is decoupled from that of ${\bf T}_{12}$ and ${\bf 1}$ in Eqs.~(30) and~(31) when $\bar{\boldsymbol{\omega}}_1$ and $\bar{\boldsymbol{\omega}}_2$ are constant vectors. Since we have argued above that it would be inappropriate to modify Eqs.~(28) and~(29) to bring the present semiclassical theory in line with our earlier theory,\cite{Manolopoulos13} and that replacing Eq.~(27) with Eq.~(34) is therefore the only way to proceed, we should at least justify this by demonstrating that Eqs.~(28) and~(29) do indeed give the correct one-electron spin dynamics when the nuclear spin motion is suppressed and $\bar{\boldsymbol{\omega}}_1$ and $\bar{\boldsymbol{\omega}}_2$ are held fixed.

This is done in Fig.~3, which compares the $xy$ components of the quantum mechanical spin correlation tensors of each electron in the radical pair with the corresponding semiclassical results, for the same model problem that was considered in Fig.~2. Asymmetric recombination was included in the dynamics in both calculations, using Eqs.~(28) and~(29) in the semiclassical case. The quantum mechanical and semiclassical results are again seen to be identical to graphical accuracy, and we have verified that this is the case for all of the other components of the two tensors and for a variety of other test cases both with and without any symmetry. This confirms that Eqs.~(28) and~(29) are correct as written, and that the modification to Eq.~(27) in Eq.~(34) is thus the only possibility that is available to us to make the present extension to $\Delta k\not=0$ consistent with our earlier semiclassical theory for $\Delta k=0$.

\begin{figure}[t]
\centering
\resizebox{0.82\columnwidth}{!} {\includegraphics{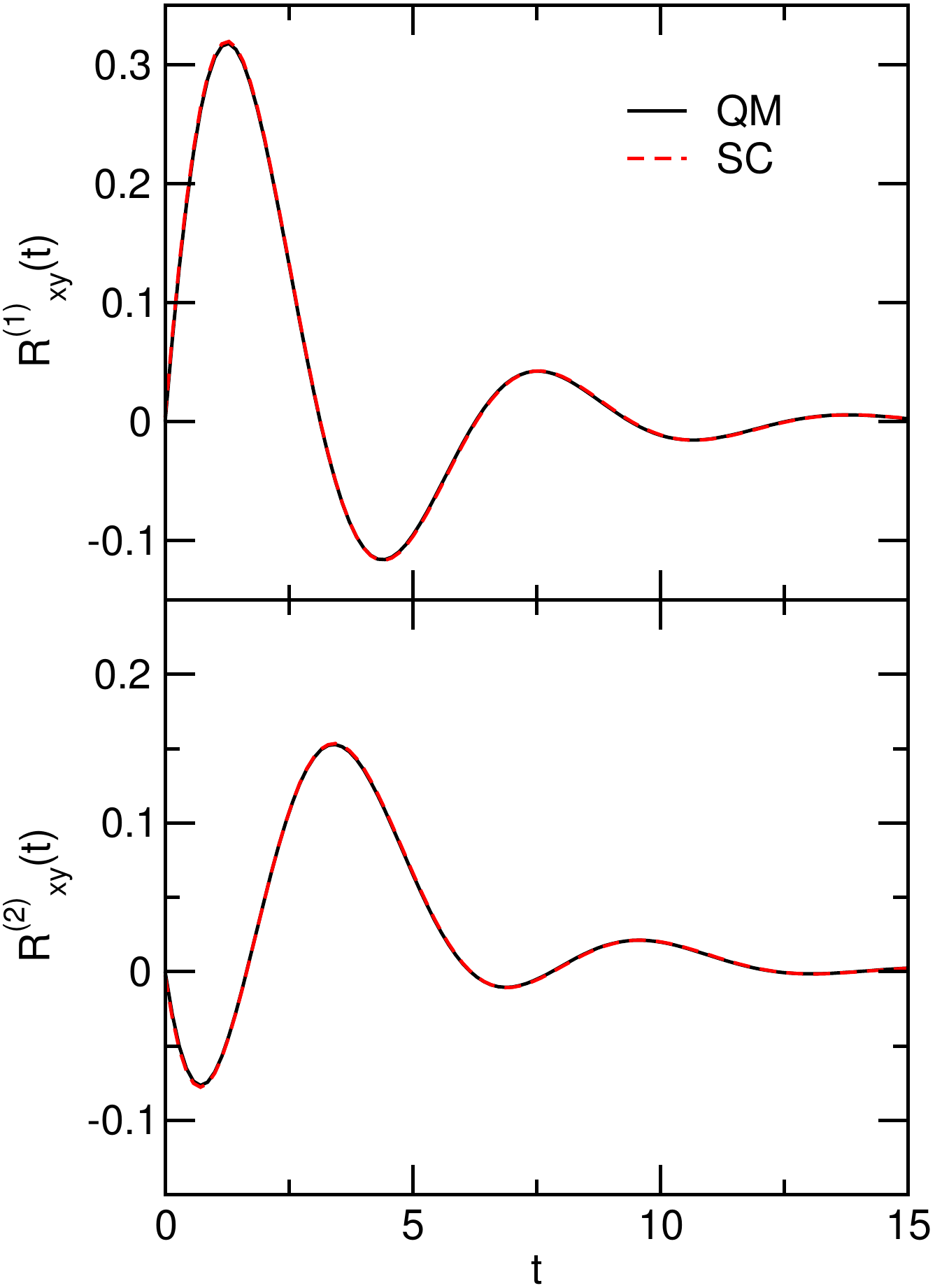}}
\caption{Comparison of the $xy$ components of the quantum mechanical (QM) and semiclassical (SC) spin correlation tensors of the electrons in radicals 1 and 2 for the asymmetric recombination problem considered in Fig.~2.}
\end{figure}

In the original semiclassical theory of Schulten and Wolynes,\cite{Schulten78} the vectors $\bar{\boldsymbol{\omega}}_1$ and $\bar{\boldsymbol{\omega}}_2$ in Eq.~(32) {\em were} held fixed during the electron spin dynamics. This can be accomplished within the present framework simply by replacing Eq.~(34) with $d{\bf I}_{ik}/dt=0$, which we shall describe in what follows as the Schulten-Wolynes approximation. As discussed in the Introduction, this approximation is inconsistent with Newton's third law of motion, because the torque exerted on the electron spins by the nuclear spins is not balanced by an equal and opposite torque on the nuclear spins from the electrons. Our earlier semiclassical theory and its present generalisation to asymmetric recombination reactions fix this problem. 

\subsection{An exact sum rule}

Finally, it is worth pointing out that the sum rule in Eq.~(17) is exactly satisfied by the semiclassical expressions for $P_{\rm S}(t)$ and $P_{\rm T}(t)$ in Eqs.~(35) and~(36), and for exactly the same reason in as in the quantum case. The crux of the argument in Eq.~(17) is the observation that
$$
k_{\rm S}\hat{P}_{\rm S}(t)+k_{\rm T}\hat{P}_{\rm T}(t) = -{d\over dt}\left(e^{+i\hat{H}t-\hat{K}t}\hat{\bf 1}e^{-i\hat{H}t-\hat{K}t}\right)
$$
$$
=-{d\over dt}\hat{\bf 1}(t), \eqno(38)
$$
which when integrated from $t=0$ to $\infty$ in the presence of recombination ($\bar{k}>0$) gives $\hat{\bf 1}(0)-\hat{\bf 1}(\infty)=\hat{\bf 1}(0)=\hat{\bf 1}$. In the semiclassical case, Eqs.~(31), (33) and (37) give
$$
k_{\rm S}\bar{P}_{\rm S}(t)+k_{\rm T}\bar{P}_{\rm T}(t) = \bar{k}{\bf 1}-4\Delta k\,{\rm tr}[{\bf T}_{12}] \phantom{xxxxxxxx}
$$
$$
= -{d\over dt}{\bf 1}(t), \eqno(39)
$$
which again integrates to give ${\bf 1}(0)-{\bf 1}(\infty)={\bf 1}(0)=1$. Therefore, from Eqs.~(33), (35) and~(36), and our prescriptions for the initial values of ${\bf 1}$ and ${\bf T}_{12}$,
$$
\Phi_{\rm S}+\Phi_{\rm T} = \prod_{i=1}^2 {2\over 4\pi}\int d\Omega_{S_i}\prod_{k=1}^{N_i}{1\over 4\pi}\int d\Omega_{I_{ik}}\, \bar{P}_{\rm S}(0)\phantom{xxx}
$$
$$
= \left({2\over 4\pi}\right)^2 \int d\Omega_{S_1}\int d\Omega_{S_2} \left({1\over 4}-{\bf S}_1\cdot{\bf S}_2\right)=1.\eqno(40)
$$

\section{Results and discussion}

\subsection{Validation for a model radical pair}

\begin{figure*}[t]
\centering
\resizebox{1.8\columnwidth}{!} {\includegraphics{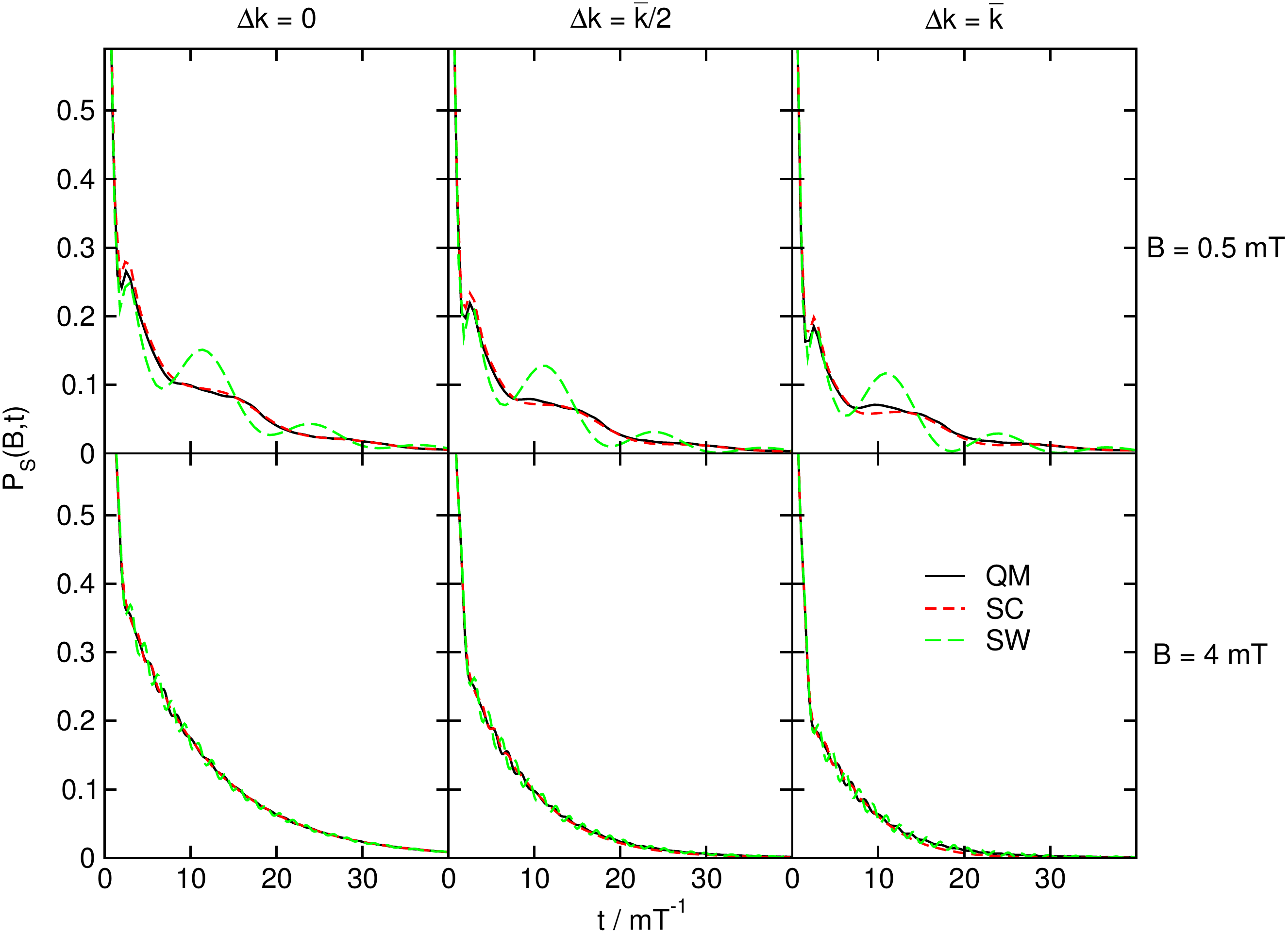}}
\caption{Comparison of quantum mechanical (QM), semiclassical (SC), and Schulten-Wolynes (SW) singlet probabilities for a model radical pair with 12 $I=1/2$ nuclear spins in one radical and none in the other, for various singlet and triplet recombination rates ($k_{\rm S}=\bar{k}+3\Delta k$ and $k_{\rm T}=\bar{k}-\Delta k$) and magnetic field strengths. $\bar{k}=0.1$ mT in all six panels.}
\end{figure*}

In order to test the above semiclassical theory, we have first applied it to a model radical pair for which the exact quantum mechanical singlet and triplet probabilities can be computed for comparison. This model, which has 12 $I=1/2$ nuclear spins in one radical and none in the other, has been used previously to show that the semiclassical theory captures the correct low magnetic field effect in the singlet yield of a radical pair reaction with symmetric recombination.\cite{Manolopoulos13} The hyperfine coupling constants of the 12 nuclear spins in the larger of the two radicals lie in the range $-1$ mT $<a_{ik}<1$ mT and are listed in Table~I of Ref.~\cite{Manolopoulos13}.

Fig.~4 compares the quantum mechanical, semiclassical, and Schulten-Wolynes singlet probabilities of this radical pair for three different recombination scenarios and two different magnetic field strengths. The recombinations considered are symmetric ($\Delta k=0$; $k_{\rm S}=k_{\rm T}=\bar{k}$), moderately asymmetric ($\Delta k=\bar{k}/2$; $k_{\rm S}=5\bar{k}/2$ and $k_{\rm T}=\bar{k}/2$), and fully asymmetric ($\Delta k=\bar{k}$; $k_{\rm S}=4\bar{k}$ and $k_{\rm T}=0$), with $\bar{k}=0.1$ mT in all three cases. The magnetic fields are 0.5 mT (in the middle of the range of $|a_{ik}|$ values where the hyperfine and Zeeman interactions are comparable), and 4 mT (approaching the high field limit where the Zeeman interaction of the electron spins with the magnetic field is beginning to dominate).

The SC results in Fig.~4 are clearly in good agreement with the exact QM singlet probabilities at both magnetic field strengths and for the full range of recombination asymmetries. Indeed there is hardly any difference between the accuracy of the SC approximation in the fully symmetric ($\Delta k=0$) and fully asymmetric ($\Delta k=\bar{k}$) cases. Note that this is the first test we have presented of the equation of motion for ${\bf I}_{ik}$ in Eq.~(34), which was not used in the construction of Figs.~2 and~3. This equation of motion, which is needed for consistency with our earlier semiclassical theory when $\Delta k=0$, clearly works equally well for $\Delta k\not=0$. By contrast, simply turning off the nuclear spin dynamics by setting $d{\bf I}_{ik}/dt=0$ does not work so well at all. This leaves an underdamped Zeeman oscillation in the resulting (SW) singlet probability with a period of $2\pi/B$ mT$^{-1}$. The nuclear spin dynamics in Eq.~(34) damps out this Zeeman oscillation and gives a singlet probability that is much closer to the correct QM result.

\begin{figure*}[t]
\centering
\resizebox{1.8\columnwidth}{!} {\includegraphics{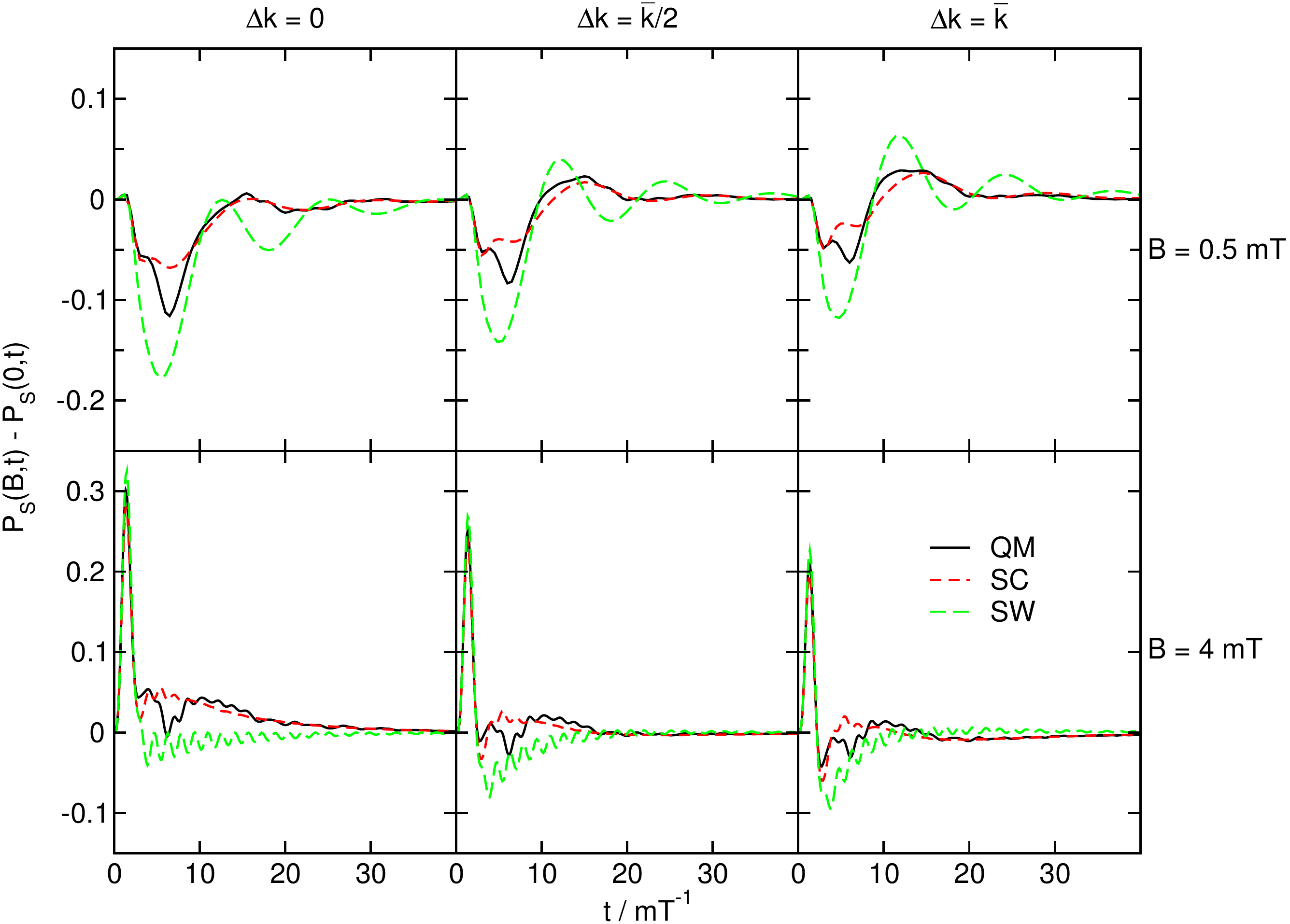}}
\caption{As in Fig.~4, but now showing the difference between the time-dependent singlet probability in the presence of a magnetic field and the same probability in the absence of a field. Note the change in the scale of the ordinate: the magnetic field effect is a small fraction of the overall signal.}
\end{figure*}

Just how much closer can be seen from Fig.~5, which shows the differences between the singlet probabilities in the presence and the absence of a magnetic field. The Zeeman oscillations in the SW results are now even more pronounced, because they make a larger relative contribution to the difference signal $P_{\rm S}(B,t)-P_{\rm S}(0,t)$ than they do to the absolute signal $P_{\rm S}(B,t)$.  Setting these Zeeman oscillations aside, the SW (fixed nuclear field) approximation is seen to be quantitatively correct at short times, but qualitatively incorrect at longer times, as evidenced by the negative tails it gives in the lower panels of Fig.~5. The SC results are qualitatively correct at both short and long times, although they too are only approximate for this $N=12$ nuclear spin problem at intermediate times. As discussed in Ref.~\cite{Manolopoulos13}, the SC theory is expected to become increasingly accurate as the number of nuclear spins increases and the environment of the electron spins becomes more complex.

\begin{figure}[t]
\centering
\resizebox{0.8\columnwidth}{!} {\includegraphics{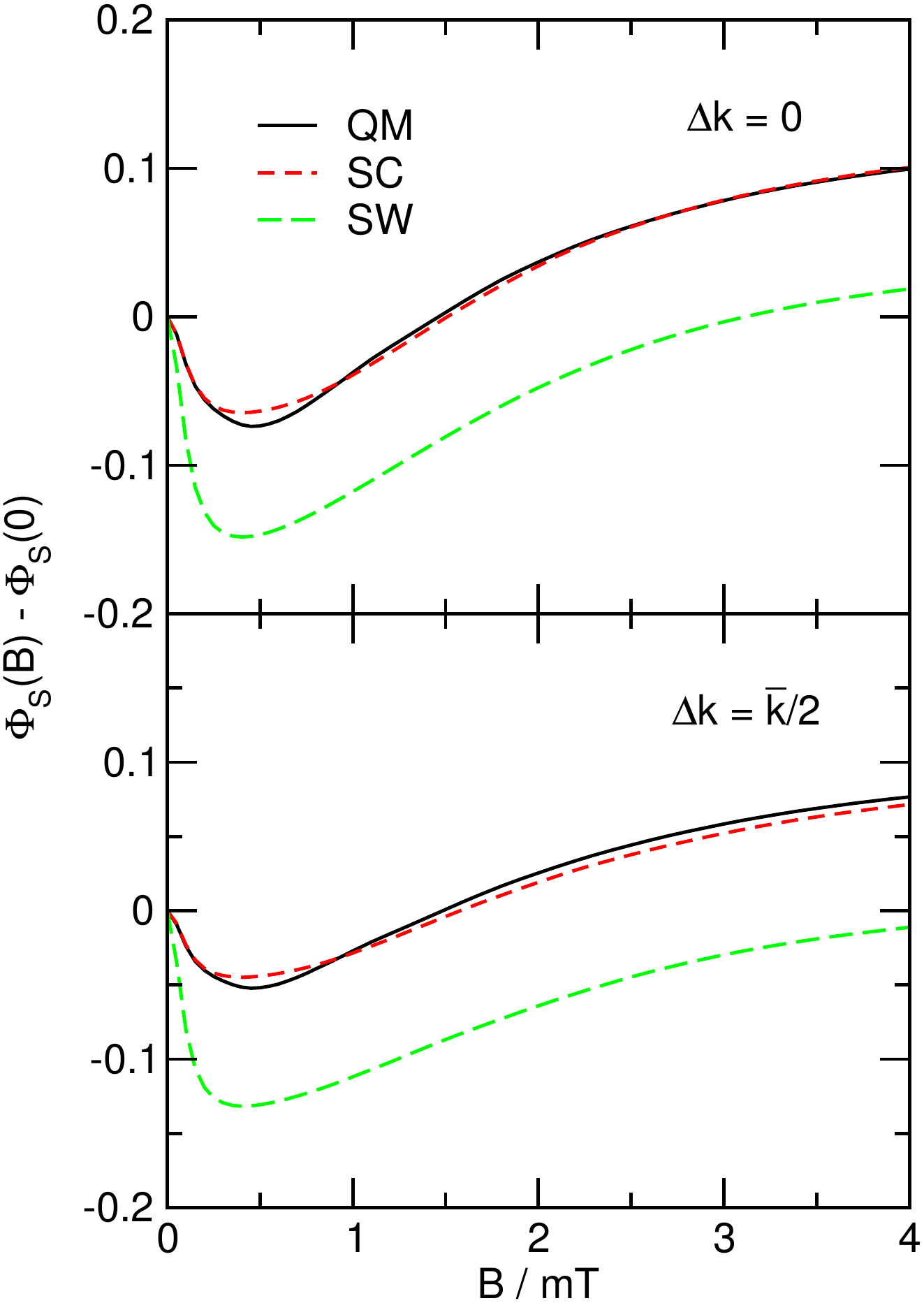}}
\caption{Comparison of QM, SC and SW low field effects on the singlet yield of a model radical pair recombination reaction with 12 $I=1/2$ nuclear spins, for both symmetric (upper panel) and asymmetric (lower panel) recombination with $\bar{k}=0.1$ mT.}
\end{figure}

The magnetic field effects in the time-dependent singlet probabilities in Fig.~5 can be converted into magnetic field effects on the singlet yield of the reaction using Eq.~(15). These singlet yield effects are shown for $\Delta k=\bar{k}$ and $\Delta{k}=\bar{k}/2$ in Fig.~6. We have not included the fully asymmetric recombination case ($\Delta k=\bar{k}$) in this figure because all three methods (QM, SC, and SW) give $\Phi_{\rm S}(B)-\Phi_{\rm S}(0)=0$ in this case by virtue of the sum rules in Eqs.~(17) and~(40). The results in the upper panel of Fig.~6 are graphically indistinguishable from those we presented in Ref.~\cite{Manolopoulos13}, which were obtained using the original version of the SC theory in Eqs.~(18)--(26). However, they were obtained here using the more general version of the theory in Eqs.~(28)--(37), which was also used to generate the lower panel of the figure. The upshot of Fig.~6 is that the present generalisation of the SC theory works equally well for the low magnetic field effect in the singlet yield of an asymmetric recombination reaction as it does for a symmetric recombination reaction, and that the fixed nuclear field (SW) approximation is not good enough to capture the correct low field effect in either case.

\begin{figure}[t]
\centering
\resizebox{0.8\columnwidth}{!} {\includegraphics{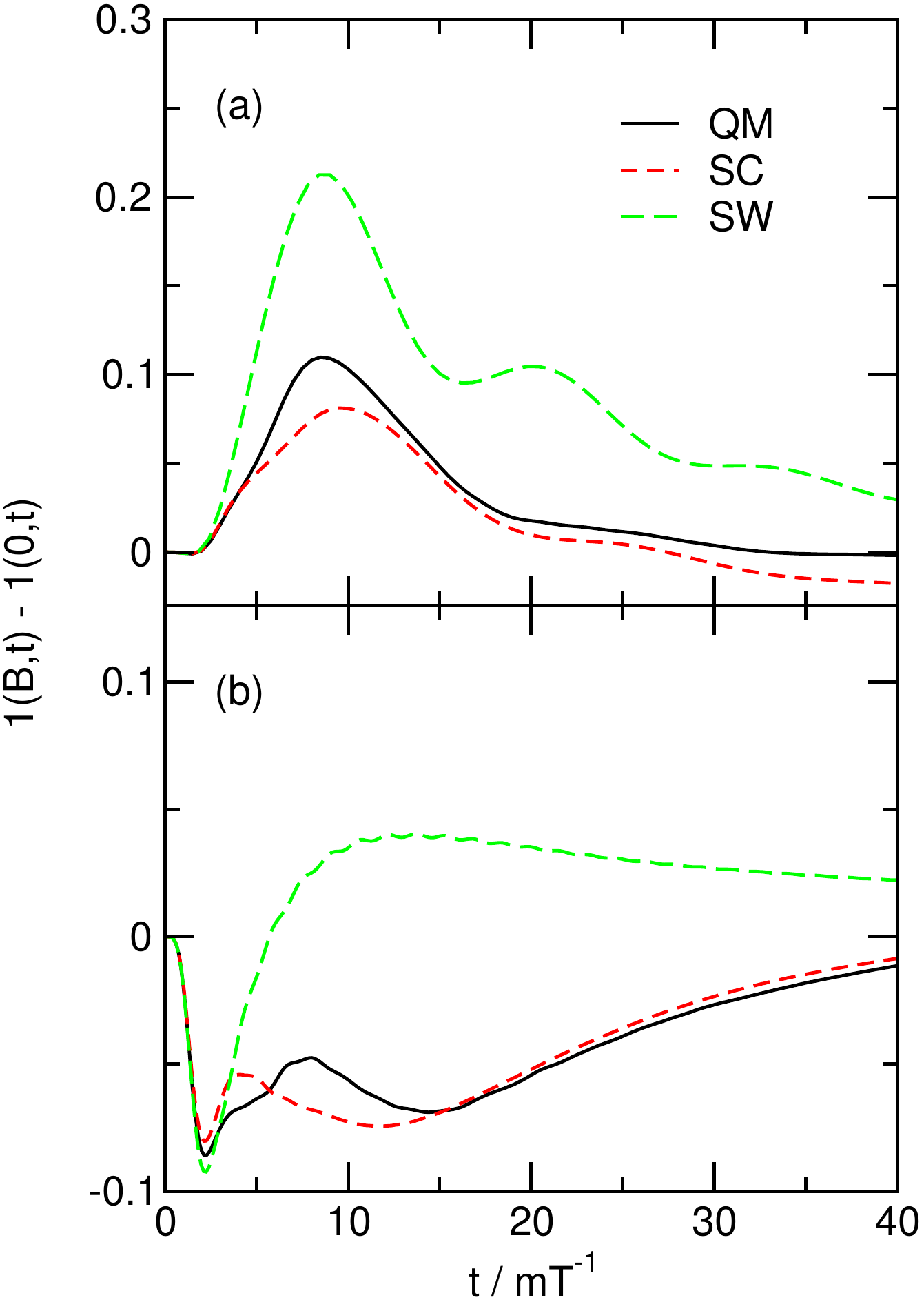}}
\caption{Comparison of QM, SC and SW magnetic field effects on the total survival probability ${\bf 1}(B,t)=P_{\rm S}(B,t)+P_{\rm T}(B,t)$ of a model radical radical pair recombination reaction with 12 $I=1/2$ nuclear spins, for the case of fully asymmetric recombination ($\Delta k=\bar{k}$). (a) $B=0.5$ mT, (b) $B=4$ mT.}
\end{figure}

Finally, since it is more closely related to the experimental observable of interest in the next subsection, we have also calculated the magnetic field effect ${\bf 1}(B,t)-{\bf 1}(0,t)$ on the total survival probability of the radical pair, ${\bf 1}(B,t)=P_{\rm S}(B,t)+P_{\rm T}(B,t)$. The results of these calculations are shown in Fig.~7 for the case of fully asymmetric recombination ($\Delta k=\bar{k}=0.1$ mT), for magnetic fields of 0.5 and 4 mT. One again sees that the SW approximation is not good enough, and that the present SC theory is almost quantitatively accurate at both short and long times. As we have already mentioned, we would expect the SC results to become more accurate for all experimental observables involving the electron spins (including this one) for a radical pair with more nuclear spins.\cite{Manolopoulos13}

\begin{figure*}[t]
\centering
\resizebox{1.6\columnwidth}{!} {\includegraphics{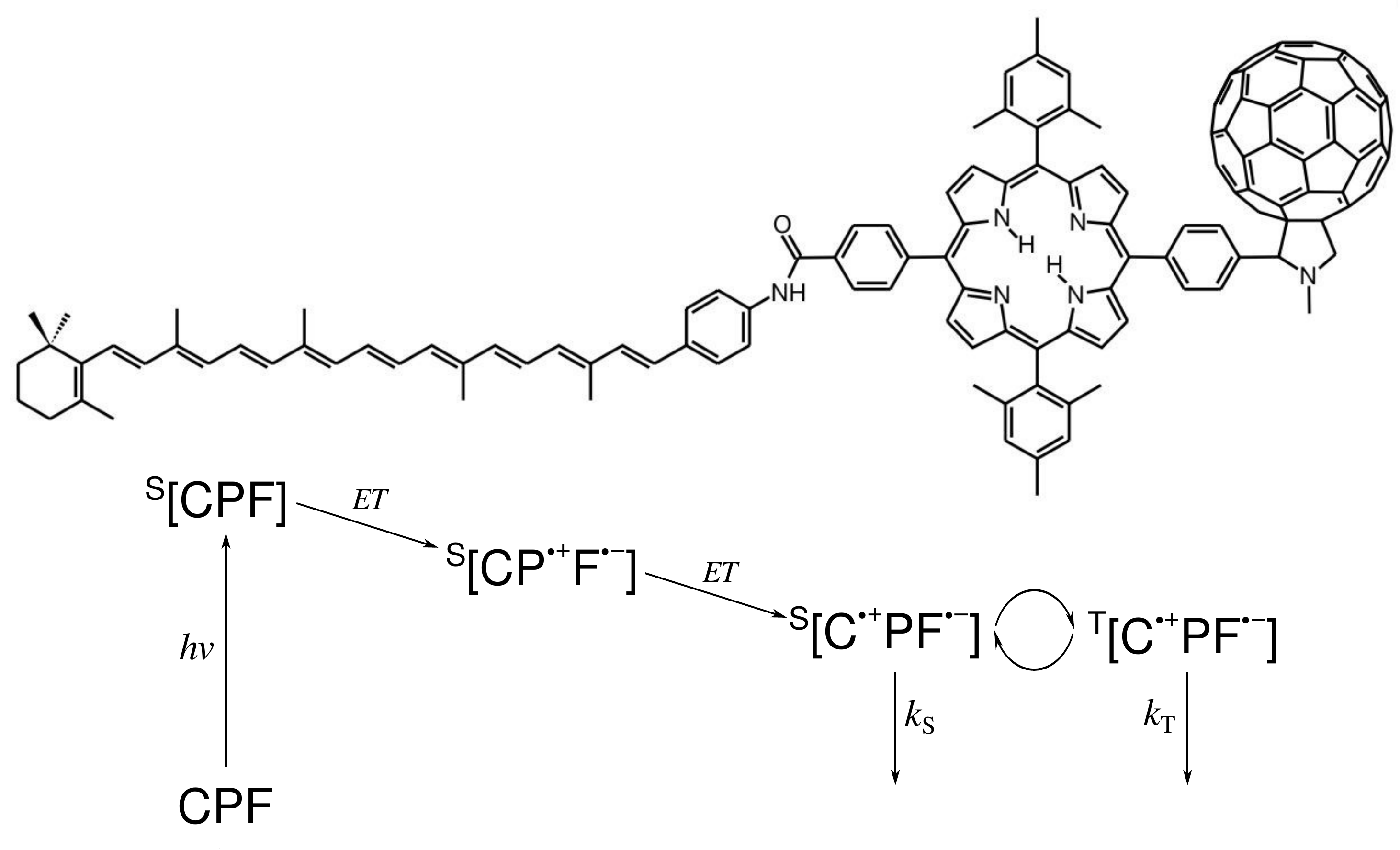}}
\caption{The carotenoid-porphyrin-fullerene triad molecule that was studied experimentally in Ref.~\cite{Maeda08}, along with an illustration of the photo-excitation and electron transfer (ET) steps that precede the coherent electron spin evolution and asymmetric recombination of the C$^{\cdot +}$PF$^{\cdot -}$ radical pair.}
\end{figure*}

\begin{figure}[t]
\centering
\resizebox{0.8\columnwidth}{!} {\includegraphics{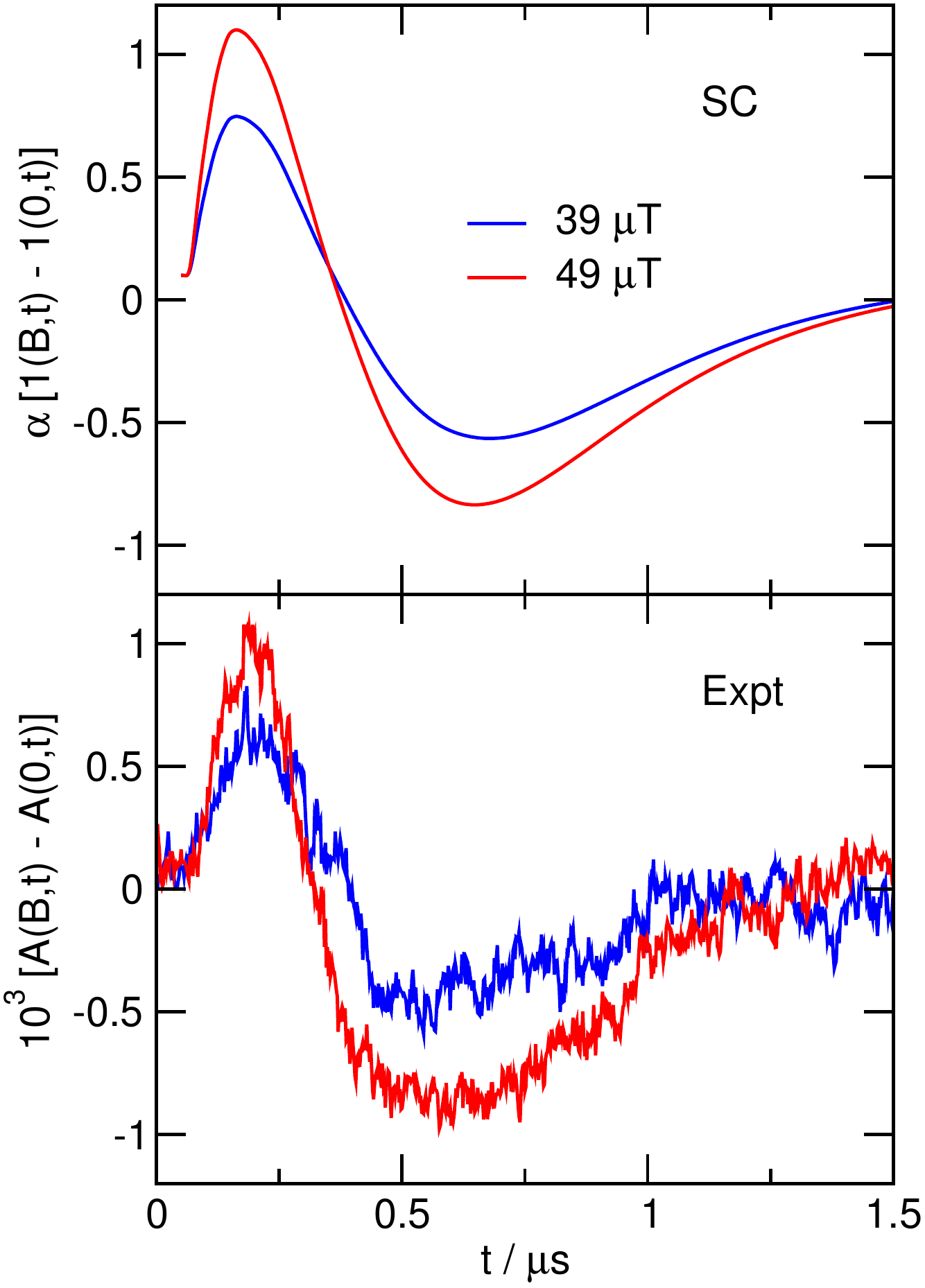}}
\caption{Upper panel: Computed SC magnetic field effects on the total survival probability of the C$^{\cdot +}$PF$^{\cdot -}$ radical pair at $B=39$ $\mu$T and 49 $\mu$T, multiplied by $\alpha=821$ to bring the positive peak in the 49 $\mu$T signal to 1, and then shifted by 0.1 to mimic the experimental background and delayed by 50 ns to mimic the experimental instrumental delay. Lower panel: Changes in the transient absorption signal of the carotenoid radical in C$^{\cdot +}$PF$^{\cdot -}$ at 113 K caused by these applied magnetic fields; experimental data from Ref.~\cite{Maeda08}}
\end{figure}

\begin{figure}[t]
\centering
\resizebox{0.8\columnwidth}{!} {\includegraphics{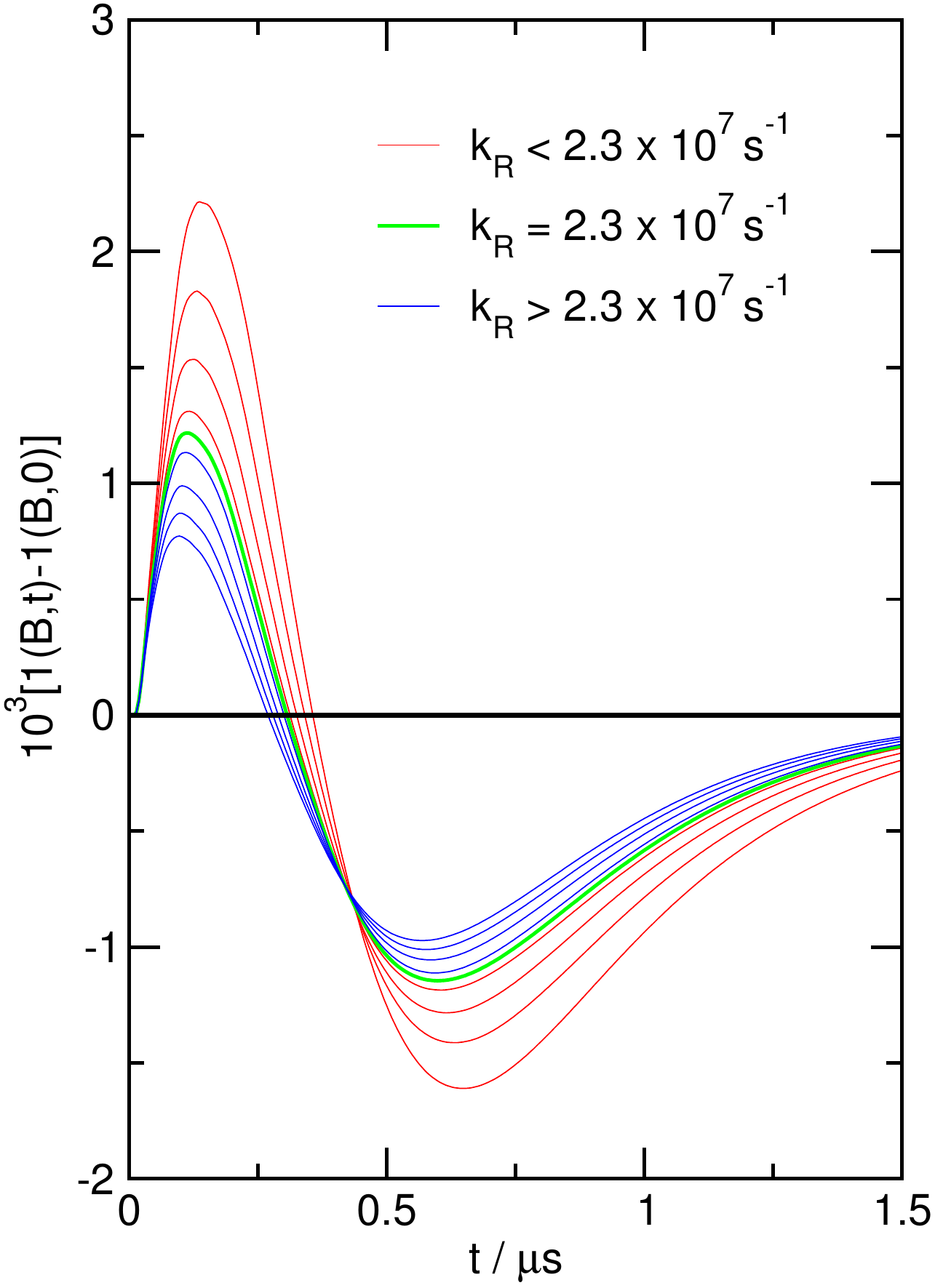}}
\caption{Dependence of the 49 $\mu$T SC results in Fig.~9 on the relaxation rate $k_{\rm R}$ of the electron spin on the carotenoid radical. Red curves: $k_{\rm R}=1.6 \times 10^7$ s$^{-1}$ to $2.2\times 10^7$ s$^{-1}$ in steps of $2\times 10^{6}$ s$^{-1}$. Green curve: $k_{\rm R}=2.3\times 10^7$ s$^{-1}$, which gives the best fit to the experimental data. Blue curves: $k_{\rm R}=2.4\times 10^{7}$ s$^{-1}$ to $3.0\times 10^{7}$ s$^{-1}$ in steps of $2\times 10^{6}$ s$^{-1}$.}
\end{figure}

\begin{figure}[t]
\centering
\resizebox{0.8\columnwidth}{!} {\includegraphics{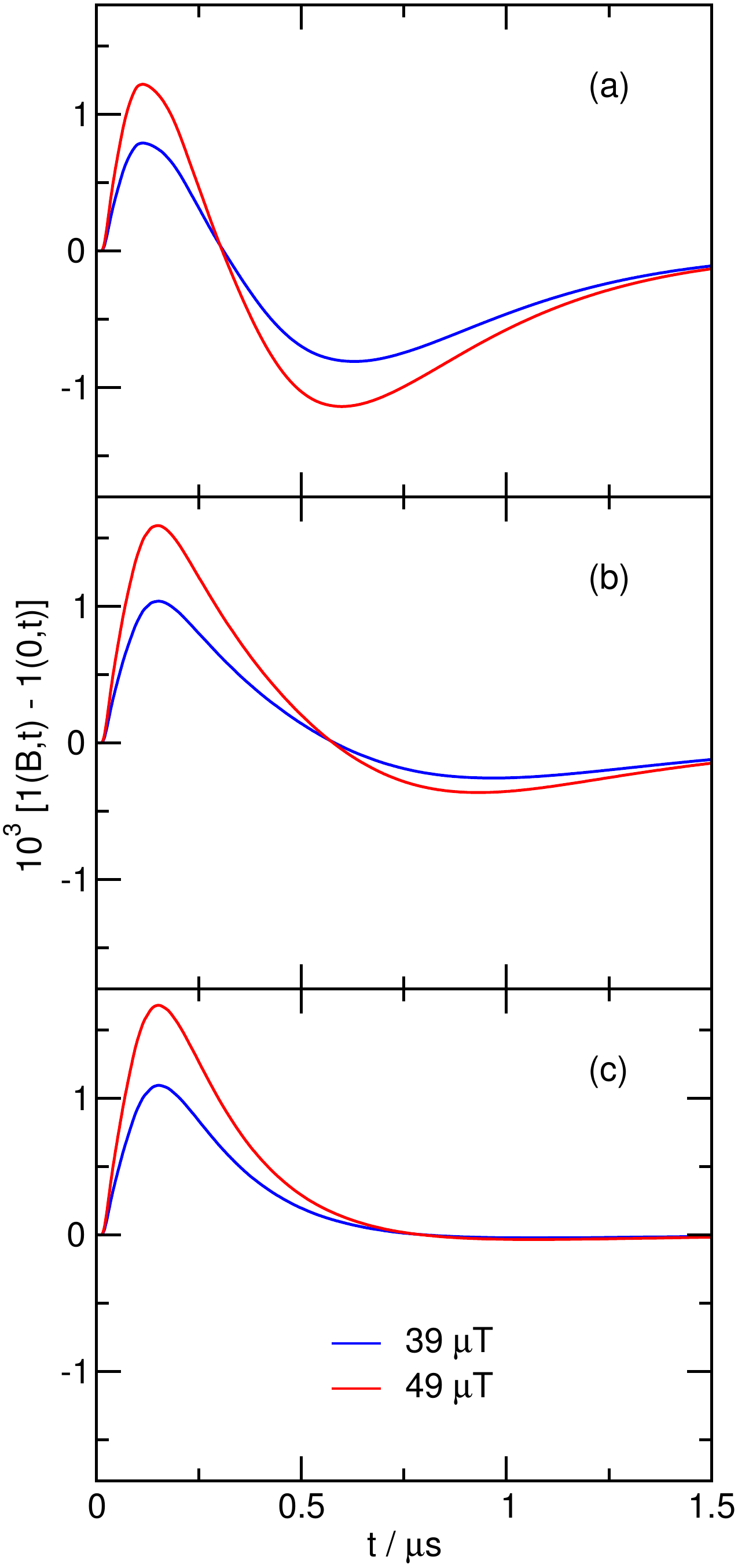}}
\caption{The effect of (artificially) moving the electron spin relaxation from the carotenoid radical (1) to the fullerene radical (2) in the C$^{\cdot +}$PF$^{\cdot -}$ radical pair. (a) Relaxation of the carotenoid electron spin: $k_{\rm R}^{(1)} = 2.3\times 10^7$ s$^{-1}$ and $k_{\rm R}^{(2)}=0$. (b) Relaxation of both electron spins: $k_{\rm R}^{(1)}=1.15\times 10^7$ s$^{-1}$ and $k_{\rm R}^{(2)}=1.15\times 10^7$ s$^{-1}$. (c) Relaxation of the fullerene electron spin: $k_{\rm R}^{(1)}=0$ and $k_{\rm R}^{(2)}=2.3\times 10^7$ s$^{-1}$.}
\end{figure}

\subsection{Application to a chemical compass}

The model radical pair with 12 nuclear spins considered above is already approaching the largest spin system that it is practical to study quantum mechanically.\cite{Manolopoulos13} However, because the computational effort in the SC method increases only linearly with the number of nuclear spins rather than exponentially, this method can be applied to arbitrarily large radicals, and used to address many interesting questions in spin chemistry. To illustrate this, we shall now present the first full-dimensional study of the spin dynamics of the carotenoid-porphyrin-fullerene (CPF) triad\cite{Kodis04} that has recently been established as a \lq proof of principle' for the operation of a chemical compass.\cite{Maeda08} 

This triad molecule is illustrated in Fig.~8, which also contains a summary of its photochemistry. Following the absorption of a visible photon by the porphyrin chromophore, there are two rapid (sub nanosecond) electron transfers, first from the excited state of the porphyrin to the fullerene (a well-established electron acceptor), and then from the carotenoid to the porphyrin.\cite{Maeda08} These electron transfers result in a C$^{\cdot+}$PF$^{\cdot-}$ radical pair which is believed to be formed predominantly in its singlet state.\cite{Maeda11} The radical pair then undergoes coherent electron spin evolution to the triplet state while the singlet and triplet components recombine at different rates, in accordance with the asymmetric recombination scheme we have been considering throughout this paper.  

A convenient experimental probe of the recombination dynamics is provided by the transient absorption signal of the carotenoid radical at time $t$ after the initial photo-excitation laser pulse.\cite{Maeda08} Since this radical is present in both the singlet and triplet states of the photo-excited triad, and disappears on recombination, its absorption is proportional to the total survival probability ${\bf 1}(B,t)=P_{\rm S}(B,t)+P_{\rm T}(B,t)$ of the radical pair. Of particular interest from the point of view of the chemical compass is how the applied magnetic field $B$ affects this survival probability, which is encapsulated in the magnetic field effect ${\bf 1}(B,t)-{\bf 1}(0,t)$. This can be accessed experimentally by measuring the transient absorption of the carotenoid radical in the presence and the absence of the applied field.\cite{Maeda08} Note that, if the recombination were symmetric, $k_{\rm S}=k_{\rm T}=\bar{k}$, one would have ${\bf 1}(B,t)=e^{-\bar{k}t}$ for all $B$, so there would be no magnetic field effect. The success of this experiment therefore rests entirely on the asymmetry of the recombination process.

In their pioneering paper, Maeda {\em et al.}\cite{Maeda08} performed a number of experiments of this form on the CPF triad, under different physical conditions and with different magnetic field strengths. By dissolving the molecule in a nematic liquid crystal so as to obtain an aligned sample that could be probed directly, and separately applying a transient absorption photoselection technique to a frozen solution of randomly oriented molecules, they were able to measure the anisotropy of the magnetic field effect on the disappearance kinetics of the radical pair, and thereby demonstrate the potential of the triad to behave as a compass.\cite{Maeda08} Unfortunately, the sensitivity of these measurements was not sufficient to detect any anisotropy in the Earth's magnetic field ($\sim 50$ $\mu$T), so higher fields ($\sim 3$ mT) were used instead. The fact that the CPF triad {\em is} capable of detecting an Earth-strength magnetic field was demonstrated in a separate transient absorption experiment, using a sample in solution at 113 K with magnetic fields of $39$ and $49$ $\mu$T.\cite{Maeda08} A number of more refined experiments are currently underway in Oxford using the same triad and related variants, but it is this original demonstration that a radical pair recombination reaction is capable of detecting the Earth's magnetic field that we shall focus on.

To use our SC theory to shed light on this experimental observation, we shall clearly need to know the isotropic hyperfine coupling constants of the nuclear spins in the C$^{\cdot +}$PF$^{\cdot -}$  radical pair, and we ought also to consider the effect of electron spin relaxation, which will almost certainly have a bearing on experimental measurements of the triad in solution. In what follows, we shall adopt the simplest possible approaches to these two problems, leaving refinements to future work. 

Inspection of Fig.~8 shows that the carotenoid in the triad contains 45 hydrogen nuclei, and that the porphyrin linker contains a number of hydrogens and nitrogens. Discounting $^{13}$C nuclei, which will be distributed randomly in the molecule with a $\sim$1\% natural abundance, these are the only magnetic nuclei that are present. The simplest approach is therefore to assume that the hyperfine interactions between the unpaired electrons in the two radicals and the magnetic nuclei on the porphyrin linker can be neglected, to treat the carotenoid radical as having 45 $I=1/2$ nuclear spins, and the fullerene radical as having none. This reduces the spin problem to a form that is identical to the model problem with 12 $I=1/2$ nuclear spins we have considered above, but with more nuclear spins than would be manageable in an exact quantum mechanical calculation.

The isotropic Fermi contact couplings of the 46 protons in the carotenoid cation ($^2$CH$^{\cdot +}$) have been calculated by Kuprov using unrestricted B3LYP\cite{Becke93,Lee88} density functional theory with a hyperfine-optimised EPR-II\cite{Barone96} orbital basis set, and we have used the 45 of these 46 coupling constants that pertain to the protons in the carotenoid radical of the CPF triad in our calculations.\cite{Footnote2} (One might question whether terminating the carotenoid cation with a covalently-bonded hydrogen atom provides a reasonable approximation to the corresponding radical in the photo-excited CPF triad. However, this closure does at least ensure that the electron hole resides entirely on the carotenoid, which is consistent with our treatment of the electron spin dynamics. When one attempts a more sophisticated closure, such as including the porphyrin linker before closing things off with a covalently-bonded hydrogen atom, one runs into the self-interaction problem of density functional theory, which leads to excessive delocalisation of the unpaired electron density onto the closing group. The B3LYP functional contains a fraction of exact exchange that reduces this problem, but there is no guarantee that it will suppress the delocalisation correctly in every situation, and in the present context it has been shown not to.\cite{Newman13} In any event, to make our spin dynamics calculations reproducible, we have listed all 45 of the hyperfine coupling constants that we actually used for the carotenoid radical in the Appendix.)

This leaves electron spin relaxation, which will clearly have an influence on measurements of the CPF triad in solution where the anisotropic (dipolar) hyperfine interactions in the carotenoid radical will be modulated by  molecular tumbling motion. To include this effect, we shall assume that the coupling between the unpaired electrons in the radical pair is sufficiently weak to exclude concerted processes, and treat the two electrons as relaxing independently. This can be done in a phenomenological way in the SC theory simply by adding the following terms to the equations of motion in Eqs.~(28)-(30):
$$
{d\over dt}{\bf S}_1 = \cdots-{\bf R}^{(1)}{\bf S}_1, \eqno(41)
$$
$$
{d\over dt}{\bf S}_2 = \cdots-{\bf R}^{(2)}{\bf S}_2, \eqno(42)
$$
$$
{d\over dt}{\bf T}_{12} = \cdots-{\bf R}^{(1)}{\bf T}_{12}-{\bf T}_{12}{\bf R}^{(2)}. \eqno(43)
$$
Here the dots ($\cdots$) denote the electron spin evolution and recombination terms that are already present in the equations and
$$
{\bf R}^{(i)} = 
\begin{pmatrix}
1/T^{(i)}_2 & 0 & 0\\
0 & 1/T^{(i)}_2 & 0\\ 
0 & 0 & 1/T^{(i)}_1
\end{pmatrix} \eqno(44)
$$ 
is a relaxation matrix for the electron spin on radical $i$. Note that these additional terms cause ${S}_{1z}$ to relax at a rate $1/T_1^{(1)}$, ${S}_{2x}$ to relax at a rate $1/T_2^{(2)}$, and $T_{1z,2x}={S}_{1z}{S}_{2x}$ to relax at a rate $1/T_1^{(1)}+1/T_2^{(2)}$. Note also that the relaxation reduces the magnitude of ${\rm tr}[{\bf T}_{12}]$, which transfers probability between $\bar{P}_{\rm S}={1\over 4}{\bf 1}-{\rm tr}[{\bf T}_{12}]$ and $\bar{P}_{\rm T}={3\over 4}{\bf 1}+{\rm tr}[{\bf T}_{12}]$ and would bring both to their equilibrium values $\bar{P}_{\rm S}={1\over 4}{\bf 1}$ and $\bar{P}_{\rm T}={3\over 4}{\bf 1}$ in the absence of any recombination. And note finally that, since the equation of motion for ${\bf 1}$ in Eq.~(31) is unchanged, Eq.~(39) continues to hold and the sum rule constraint in Eq.~(40) is unaffected by the relaxation. All of this is physically correct.

The inclusion of electron spin relaxation using Eqs.~(41)--(44)  introduces four new parameters, the longitudinal ($T^{(i)}_1$) and transverse ($T^{(i)}_2$) relaxation times of the electrons in the two radicals. However, for the CPF triad in an Earth-strength magnetic field, we can use the physics of the problem to reduce these four parameters to just one. Since the Earth's magnetic field is so weak, the electron spin relaxation along the field ($z$) axis will be hardly any different from that along the $x$ and $y$ axes, so we can make the approximation that $T^{(i)}_1\simeq T^{(i)}_2$. And since the mechanism of the electron spin relaxation is the modulation of its hyperfine interactions by the tumbling motion in solution, and the fullerene radical (let us call it 2) in the triad has no hyperfine interactions, we can make the further approximation that $1/T^{(2)}_1\simeq 1/T^{(2)}_2\simeq 0$. With these simplifications, the only remaining parameter is the relaxation rate $k_{\rm R}=1/T_1^{(1)}=1/T_2^{(1)}$ of the electron spin on the carotenoid radical.

Moreover this is the {\em only} adjustable parameter that we shall need to include in our calculations. We shall also clearly need to know the singlet and triplet recombination rates $k_{\rm S}$ and $k_{\rm T}$ of the radical pair, but these have been determined experimentally.\cite{Maeda11} By fitting a simple kinetic model to the decays recorded in pulsed electron paramagnetic resonance (EPR) measurements of the triad at 110 K, Maeda {\em et al.}\cite{Maeda11} have established that $k_{\rm S}\simeq 1.8\times 10^{7}$ s$^{-1}$ and $k_{\rm T}\simeq 7.1\times 10^{4}$ s$^{-1}$, and that the fraction of the triplet in the initial state of the C$^{\cdot +}$PF$^{\cdot -}$ radical pair is $\lambda\simeq 0.07$. In what follows, we shall take all three of these numbers as given, and assume that they apply equally well to the earlier experiment in an Earth-strength magnetic field at 113 K.\cite{Maeda08} The small initial fraction of the triplet state in the radical pair can be included in the SC theory by replacing $\bar{P}_{\rm S}(0)$ in Eqs.~(35) and~(36) with
$$
\bar{P}_{\rm S}(0)\to (1-\lambda)\bar{P}_{\rm S}(0)+(\lambda/3)\bar{P}_{\rm T}(0), \eqno(45)
$$
which populates all three components of the triplet equally at time $t=0$. This is consistent with the kinetic model that Maeda {\em et al.} used to interpret their pulsed EPR data.\cite{Maeda11}

With all of these preliminaries in hand, we are finally in a position to compare the computed SC magnetic field effect on the survival probability of the C$^{\cdot +}$PF$^{\cdot -}$ radical pair with the experimental measurements of Maeda {\em et al.}\cite{Maeda08} This comparison is shown for an optimised carotenoid radical relaxation rate of $k_{\rm R}=2.3\times 10^7$ s$^{-1}$ in Fig.~9. By tuning just this one free parameter, one sees that the SC theory can be brought into almost quantitative agreement with the experiment. The  experimental signal is reproduced equally well at both field strengths, both in terms of its time scale (which was obtained using the gyromagnetic ratio of a free electron in the SC calculations, $\gamma_{\rm e}=176$ $\mu$s$^{-1}$ mT$^{-1}$) and of the relative amplitudes of its positive and negative peaks. Furthermore, the magnitude of the SC magnetic field effect is also roughly correct. In the 49 $\mu$T experimental data, the difference signal $A(B,t)-A(0,t)$ is $\sim 1.5$\% of the field-free carotenoid absorption signal $A(0,t)$ at 0.7 $\mu$s,\cite{Maeda08} and in our SC calculations ${\bf 1}(B,t)-{\bf 1}(B,0)$ is $\sim 1.5$\% of ${\bf 1}(0,t)$ at 0.65 $\mu$s, which allows for the $\sim 50$ ns experimental instrumental delay.

The dependence of the SC results on the relaxation rate $k_{\rm R}$ of the electron spin on the carotenoid radical is illustrated for the case of a 49 $\mu$T applied field in Fig.~10. The magnitude and the time scale of the computed magnetic field effect both decrease as the electron spin relaxation rate increases, as one would expect. This monotonic behaviour makes it straightforward to determine the $k_{\rm R}$ that gives the best fit to the experimental data, and the sensitivity of the results to $k_{\rm R}$ fixes the optimum value quite precisely: neither $k_{\rm R}=2.2\times 10^{7}$ s$^{-1}$ nor $k_{\rm R}=2.4\times 10^{7}$ s$^{-1}$ gives quite such good agreement with the experimental signals at both magnetic field strengths as the SC calculation with $k_{\rm R}=2.3\times 10^{7}$ s$^{-1}$. 

The most intriguing aspect of the experimental results is the biphasic nature of the time signal. This has been the subject of some debate. An exotic early explanation based on quantum measurement theory invoked the quantum Zeno effect under the assumption that $k_{\rm T}\gg k_{\rm S}$.\cite{Kominis09} However, this has since been ruled out by the pulsed EPR measurements of Maeda {\em et al.},\cite{Maeda11} which clearly show that $k_{\rm S}\gg k_{\rm T}$. In fact, it is easy to see from this (and the predominantly singlet initial state) why a weak magnetic field enhances the survival probability of the radical pair at short times. This is the standard low field effect:\cite{Brocklehurst76,Timmel98,Brocklehurst96,Till98} a small applied field breaks the symmetry of the zero-field problem and splits the degeneracy of the zero-field eigenstates, producing more pathways for singlet to triplet interconversion. Applying the field therefore converts more of the predominantly singlet initial state into the triplet state, where it recombines more slowly when $k_{\rm S}\gg k_{\rm T}$ thereby increasing the survival probability of the radical pair. 

What is harder to understand is why the applied field decreases the survival probability at longer times. Maeda {\em et al.}\cite{Maeda11} have provided an explanation for this based on a combination of the small initial population of the triplet state and relaxation-induced population flow between the T$_{\pm 1}$ triplet components and the S and T$_0$ states in the high-field limit, which they backed up with a simplified quantum mechanical calculation involving just two effective hyperfine interactions in a magnetic field of 1.28 mT. However, the experimental results in Fig.~9 are far from the high-field limit, and the present SC calculations which reproduce these results suggest that the low-field situation is significantly more complicated. 

The most compelling evidence we have found for this is presented in Fig.~11. If it were true that the biphasic behaviour of the experimental signal resulted solely from relaxation-induced population flow between the T$_{\pm 1}$ and S and T$_0$ states, one would not expect it to make any difference whether the relaxation were of the electron spin on the carotenoid, the electron spin on the fullerene, or both (the last of these possibilities being the one considered in Maeda {\em et al.}'s calculation\cite{Maeda11}). The population transfer between the singlet and triplet states of the radical pair operates in the same way regardless of which electron spin it is that is relaxing.

Fig.~11 shows that the SC results are however strongly dependent on the details of the electron spin relaxation. The negative peak in ${\bf 1}(B,t)-{\bf 1}(0,t)$ that is seen when the relaxation is confined to the carotenoid electron is significantly damped when both electrons relax at half the rate, and washed out almost completely when all of the relaxation is transferred to the fullerene electron. Yet the equation of motion for ${\bf T}_{12}$ in Eq.~(43), which governs the transfer of population between $\bar{P}_{\rm S}={1\over 4}{\bf 1}-{\rm tr}[{\bf T}_{12}]$ and $\bar{P}_{\rm T}={3\over 4}{\bf 1}+{\rm tr}[{\bf T}_{12}]$, is  exactly the same in all three cases. What does change is the dynamics of the individual electron spins, and evidently the biphasic behaviour is only captured correctly (with $k_{\rm R}=2.3\times 10^{7}$ s$^{-1}$) when it is the electron spin on the carotenoid that relaxes. (The biphasic behaviour can be restored in the other two scenarios by reducing $k_{\rm R}$, but then its time scale becomes too long to match the experiment.) The upshot is that the SC simulation can only be brought into agreement with the experiment in the physically realistic scenario in which it is the electron spin on the carotenoid radical that relaxes (recall that discounting any $^{13}$C's that may be present the fullerene radical does not have any nuclear spins available to relax its electron spin via dipolar coupling).

We conclude that the biphasic behaviour of the experimental signal arises from a delicate balance between the asymmetric recombination of the radical pair (with $k_{\rm S}=1.8\times 10^7$ s$^{-1}\gg k_{\rm T}$) and the relaxation of the electron spin on the carotenoid radical (with $k_{\rm R}=2.3\times 10^7$ s$^{-1}$). This balance is  too subtle to be captured correctly by a back-of-the-envelope calculation, at least in an Earth-strength magnetic field where the Zeeman and hyperfine interactions in the carotenoid radical are comparable and there is no obvious simplification that can be made to its spin dynamics. However, it has emerged as the only plausible way to explain the experimental data from our SC calculations, which we would expect on the basis of Figs.~4-7 to be quite reliable for a radical pair of the size of the C$^{\cdot +}$PF$^{\cdot -}$ triad.

The only remaining question is whether the relaxation rate of $k_{\rm R}=2.3\times 10^7$ s$^{-1}$ that we have obtained for the electron spin on the carotenoid radical is reasonable. In their fit to the effect of a larger (1.28 mT) magnetic field on the survival probability of the same triad at 110 K, which used a simplified model with just two effective hyperfine interactions and assumed equal relaxation rates for both electron spins, Maeda {\em et al.}\cite{Maeda11} obtained a value of $k_{\rm R}=2.1\times 10^6$ s$^{-1}$, a full order of magnitude smaller than our relaxation rate. However, since this was for two electron spins rather than one, we should double it for a fairer comparison of its effect on the rate of transfer of population between the singlet and triplet states of the radical pair. Then the difference is only a factor of 5. Given that the experimental data used in the two fits was measured at different magnetic field strengths ($<50$ $\mu$T and 1.28 mT) and temperatures (113 K and 110 K), that both of these temperatures are very close to the $\sim 110$ K melting point of the 2-methyltetrahydrofuran solvent that was used in the experiments, and that Maeda {\em et al.} were able to use a far cruder theoretical model to fit their high-field data, we do not find this factor of 5 at all unreasonable.

\section{Conclusions}

In this paper, we have generalised our earlier SC theory of radical pair recombination reactions\cite{Manolopoulos13} to allow for asymmetric recombination of the radical pair and relaxation of its electron spins. The resulting theory has been validated by comparison with exact QM results for a model radical pair containing 12 nuclear spins (see Figs.~4 to~7), and then applied to a carotenoid-porphyrin-fullerene triad for which the exact QM calculation would be quite impractical (see Figs.~8 to~11). This application has shed light on the response of the photo-excited C$^{\cdot +}$PF$^{\cdot -}$ triad to an Earth-strength magnetic field, suggesting in particular that the experimentally observed response arises from a delicate balance between the asymmetric recombination of the radical pair and the relaxation of the electron spin on the carotenoid radical. 

This clearly establishes the SC theory as a practical way to solve problems in spin chemistry. Before it can be used more widely, however, we shall first have to generalise it to include a number of other effects that are often relevant, such as anisotropic hyperfine interactions, exchange and dipolar couplings between the two electrons in the radical pair, and interactions with radiofrequency radiation. The present extension of the theory to the two-electron spin variables in Eqs.~(28) to~(37) paves the way for these additional generalisations, which we expect to be comparatively straightforward. Once they have been incorporated, the SC theory promises to provide a universal tool for analysing spin chemistry problems that are too complicated to study quantum mechanically.

\begin{acknowledgements}
We would like to thank Christiane Timmel for helpful discussions, Ilya Kuprov for providing us with the hyperfine coupling constants of the carotenoid radical, and Martin Galpin for spotting the key step in the argument in Eq.~(17). This work was supported by the Defence Advanced Research Projects Agency (QuBE: N66001-10-1-4061) and the European Research Council (PJH), and by the Wolfson Foundation and the Royal Society (DEM).\\
\end{acknowledgements}

\section{Appendix: Carotenoid hyperfine coupling constants}

The isotropic hyperfine coupling constants of the 45 protons in the carotenoid radical that we used in our spin dynamics calculations were as follows (in mT):  0.04879, 0.046328, $-$0.115098, $-$0.111317, $-$0.361254, 0.130081, 0.094903, $-$0.316911, 0.094676, $-$0.021817, $-$0.140593, $-$0.087963, $-$0.071456, 0.050581, $-$0.275215, 0.056448, 0.111917, $-$0.385563, 0.480492, 0.015702, 0.490845, 0.332346, 0.306659, 0.011857, 0.008375, 0.246503, 0.257005, 0.015707, 0.49151, 0.407741, 0.17369, 0.579152, 0.057321, 0.006161, $-$0.005099, $-$0.003271, 0.018443, 0.001563, $-$0.017735, 0.014287, $-$0.028314, 0.003183, 0.246942, 0.442049, 0.027488. 

Note that 17 of these hyperfine coupling constants are smaller in absolute value and the remaining 28 larger than the Earth's magnetic field ($\sim 50$ $\mu$T). The positions of the 45 protons in the radical are immaterial to the spin dynamics calculation and so are not needed to reproduce our results.


\begin{thebibliography}{99}
\bibitem{Footnote1}
The proof that the present treatment of the electron spin dynamics is exact for all $t$ for fixed fields $\bar{\boldsymbol{\omega}}_1$ and $\bar{\boldsymbol{\omega}}_2$ requires a lengthy digression into the properties of spin coherent states and their relationship to the classical vector model. Since this is an interesting topic in its own right and is peripheral to the present development which we shall defer it to a separate publication.
\bibitem{Footnote2} 
The carotenoid radical ends just before the amide linker in Fig.~8 and the extra H is attached to the phenyl ring at this point.
\end{thebibliography}
\end{document}